\documentclass[structabstract]{aa}  
\usepackage{natbib}
\usepackage{lscape}
\usepackage{rotating}
\usepackage{graphicx}
\usepackage{enumerate}
\usepackage{txfonts}
\begin{document}
  \title{Evolution towards and beyond accretion-induced collapse of\\
         massive white dwarfs and formation of millisecond pulsars}

   \author{T. M. Tauris
          \inst{1,2}\fnmsep\thanks{e-mail: tauris@astro.uni-bonn.de},
          D. Sanyal\inst{1,3},
          S.-C. Yoon\inst{1,4},
          \and
          N. Langer\inst{1}
          }

   \authorrunning{Tauris, Sanyal, Yoon \& Langer}
   \titlerunning{Evolution leading to AIC and formation of MSPs}

   \institute{Argelander-Insitut f\"ur Astronomie, Universit\"at Bonn, Auf
              dem H\"ugel 71, 53121 Bonn, Germany
         \and
              Max-Planck-Institut f\"ur Radioastronomie, 
              Auf dem H\"ugel 69, 53121 Bonn, Germany 
         \and
              Department of Physical Sciences, IISER Kolkata, 
              Mohanpur - 741252, Nadia, West Bengal, India
         \and
              Department of Physics \& Astronomy, Seoul National University,
              Gwanak-gu, Gwanak-ro 1, $\;$Seoul , 151-742, $\;$Korea
             }

   \date{Received April 8, 2013; Accepted July 30, 2013}

 
  \abstract
   {Millisecond pulsars (MSPs) are generally believed to be old neutron stars (NSs), formed via 
    type~Ib/c core-collapse supernovae (SNe), which have been spun up
    to high rotation rates via accretion from a companion star
    in a low-mass X-ray binary (LMXB). In an alternative formation channel, NSs are produced via the accretion-induced
    collapse (AIC) of a massive white dwarf (WD) in a close binary.} 
   {Here we investigate binary evolution leading to AIC 
    and examine if NSs formed in this way can subsequently be recycled to form MSPs and, if so,
    how they can observationally be distinguished from pulsars formed via the standard  
    core-collapse SN channel in terms of their masses, spins, orbital periods and space velocities.}
   {Numerical calculations with a detailed stellar evolution code were used for the first time 
    to study the combined pre- and post-AIC evolution of close binaries.
    We investigated the mass transfer onto a massive WD (treated as a point mass) in 240 systems with 
    three different types of non-degenerate donor stars: main-sequence stars, red~giants, and helium stars.
    When the WD is able to accrete sufficient mass 
    (depending on the mass-transfer rate and the duration of the accretion phase)
    we assumed it collapses to form a NS and we studied the dynamical effects of this
    implosion on the binary orbit. 
    Subsequently, we followed the mass-transfer epoch which resumes once the donor star refills its Roche~lobe 
    and calculated the continued LMXB evolution until the end.}
   {We show that recycled pulsars may form via AIC from all three types of progenitor systems investigated and 
    find that the final properties of the resulting MSPs are, in general, remarkably similar 
    to those of MSPs formed via the standard core-collapse SN channel.
    However, as a consequence of the finetuned mass-transfer rate necessary to make the WD grow in mass, the
    resultant MSPs created via the AIC~channel preferentially form in certain orbital period intervals.
    In addition, their predicted small space velocities can also be used to identify them observationally.  
    The production time of NSs formed via AIC can exceed 10~Gyr which can therefore 
    explain the existence of relatively young NSs in globular clusters.
    Our calculations are also applicable to progenitor binaries of SNe~Ia under certain conditions.
    }
   {}


   \maketitle
%

\section{Introduction}\label{sec:intro}
The final outcome of close binary stellar evolution is a pair of compact objects if the system avoids disruption following
a supernova (SN) explosion or a merger event in a common envelope (CE).
The nature of the compact objects formed can be either black holes, neutron stars (NSs), or white dwarfs (WDs), 
depending primarily on the initial stellar masses and orbital period of the 
zero-age main-sequence (ZAMS) binary.
Systems in which the primary star is not massive enough to end its life as a NS may  
leave behind an oxygen-neon-magnesium white dwarf (ONeMg~WD) of mass $M_{\rm WD}\simeq 1.1-1.3\;M_{\odot}$, which  
may stem from primaries with initial masses in the interval of $M_{\rm ZAMS}\simeq 6-16\;M_{\odot}$,
depending on the mass of the secondary star and, in particular, on the orbital period which affects
both the important carbon/oxygen--ratio at the depletion of core helium burning and the occurrence of a second dredge-up phase 
at the beginning of the asymptotic giant branch \citep[see e.g.][]{wl99,plp+04}.
In systems where an ONeMg~WD forms, the WD may later accrete sufficient material, 
when the secondary star subsequently evolves and fills its Roche~lobe, such that 
it reaches the Chandrasekhar-mass limit and implodes via an accretion-induced collapse (AIC) to form a NS
\citep[e.g.][]{nmsy79,tv86,mic87,cil90,nk91}. The donor star in these systems can either be
a main-sequence star, a low-mass giant, or a helium star.
The first aim of this paper is to investigate in which binaries AIC can occur.

Millisecond pulsars (MSPs) are traditionally believed to be old NSs which have been spun up 
to high rotation rates via accretion of mass and angular momentum from a companion star in a low-mass X-ray binary, LMXB
\citep[e.g.][and references therein]{bv91,tv06}.
It is important to investigate whether or not this standard recycling scenario is the sole formation channel of MSPs.
Three key questions arise in the context of AIC: 
\begin{enumerate}
 \item Could the implosion of the ONeMg~WD lead directly to the formation of an MSP? 
 \item When an additional accretion phase is necessary for NSs formed via AIC in order to explain the observed rapid spins and low B-fields of MSPs,
       is it possible at all to form MSPs via post-AIC accretion from the same donor star? 
 \item If so, can MSPs produced this way be distinguished observationally from MSPs that formed via the standard core-collapse SN~channel? 
\end{enumerate}

In the context of direct and indirect formation of MSPs via AIC (i.e. prompt or after additional accretion, respectively), 
we point out that the origin of pulsar B-fields is not well understood. 
In direct MSP formation, B-fields could be created from conservation of magnetic flux of the collapsing core, 
as originally hypothesized for NSs by \citet{wol64}, or as suggested by 
theoretical work on thermomagnetic effects during or shortly after the NS is formed \citep{rei03,spr08}.
In principle, a Chandrasekhar~mass WD with an equatorial radius of about 3000~km and 
a B-field of $10^3\;{\rm G}$ could, assuming flux conservation, undergo AIC and directly produce an MSP with $B\simeq 10^8\;{\rm G}$
and a spin period of a few~ms, equivalent to typical values of observed recycled radio MSPs \citep{tv06} 
and accreting X-ray MSPs \citep{pw12} near the spin-up line
in the $P\dot{P}$--diagram. Producing the spin of the MSP via AIC is not a problem if  
just a small fraction of the spin angular momentum of the WD is conserved during the collapse \citep{dbo+06}.

Until recently, it was thought that the distribution of WD B-fields is strongly bimodal, with a large majority of WDs being non-magnetic
and a smaller fraction ($\sim\!15$\%, primarily in binaries) having larger fields of typically $\sim\!10^7\;{\rm G}$ \citep{wf00,lbh03}.
However, by using more sensitive instruments it has been demonstrated 
that some $15-20$\% of WDs could have weak B-fields of the order of $10^3\;{\rm G}$ \citep[e.g.][]{jan+07}. 
Hence, AIC may lead to the formation of NSs with a potential large range of possible B-field values, if flux conservation is at work. 
One the other hand, regardless of its formation mechanism any newborn NS is extremely hot and liquid, and therefore differentially
rotating, which may amplify any seed magnetic field such that the B-fields of NSs formed via AIC could be similar to those
formed via an iron-core collapse.
In this respect, it is important that none of the more than 40 known young NSs associated with SN remnants
are observed with the characteristic properties of MSPs: low B-fields and fast spin.
The B-fields of MSPs ($10^{7-9}\;G$) are typically lower than that of young, normal pulsars ($10^{12-14}\;G$) by five orders of magnitude.
It is therefore questionable if an MSP would form directly from any type of collapse: iron-core collapse, electron capture SN, or AIC\footnote{
Although the argument for the AIC is less certain since it may not give rise to the formation of an observable remnant.}.
For the rest of this work, we therefore focus our attention exclusively on the possibility of indirect formation of MSPs via AIC, i.e.
following the scenario where an AIC leads to the formation of a normal NS which then subsequently accretes matter, resulting in 
a weaker B-field and a faster spin.

The AIC route of forming MSPs has three main advantages: 
1) it can explain the low space velocities of many recycled pulsars and the large fraction of NSs retained in globular clusters \citep{bg90} due
to both the small amount of mass lost and the small momentum kick expected to be associated with the implosion 
\citep[for details of simulations of the implosion, see e.g.][]{kjh06,dbo+06}; 
2) it can explain the existence of apparently {\it young} NSs in globular clusters \citep{lmd96,blt+11}; and finally 
3) it may explain the presence of a number of peculiar high B-field and 
slowly spinning Galactic disk NSs in close binaries with (semi)degenerate companions \citep[e.g.][]{ynv02}.
In addition, direct formation of MSPs via AIC could possibly help explain the postulated 
birthrate problem \citep{kn88} between the small number of LMXB progenitor
systems and the large observed number of MSPs.\\
A recent, detailed population synthesis study by \citet{htw+10} 
concluded that one cannot ignore the AIC route to MSP formation and that some binary MSPs in wide orbits
are best explained by an AIC scenario 
\citep[see also ][for a specialized study on NS formation in globular clusters]{ihr+08}. 
There are, however, many uncertainties involved in even the best population synthesis studies and in particular in 
the applied physical conditions for making the ONeMg~WD mass grow sufficiently.

The weaknesses of the AIC formation channel are that it lacks direct observational evidence of the AIC event itself
and, as already mentioned, the difficulty in predicting the spin rate and the surface B-field 
associated with a newborn NS formed via AIC 
\citep[e.g.][]{kjh06,dbo+06,dblo07}. 
It is not clear whether or not the AIC would be an observable transient event. 
According to \citet{dbo+06}, during the AIC only a few $0.001\;M_{\odot}$ of material is ejected (of which $\sim\!25\%$ is $^{56}{\rm Ni}$,  
decaying into $^{56}{\rm Fe}$ via $^{56}{\rm CO}$) which quickly becomes optically thin. Hence, AIC events are
most likely underluminous and very short-lived. 
The studies by \citet{kjh06,mpq09,dmq+10} also yield somewhat small amounts
($<0.015\;M_{\odot}$) of $^{56}{\rm Ni}$ ejected in the AIC process, 
possibly synthesized in a disk, which may result in a radioactively powered 
SN-like transient that peaks after $\le 1\;{\rm day}$ with a 
bolometric luminosity $\simeq 10^{41}\;{\rm erg}\;{\rm s}^{-1}$.
It is also possible that a transient radio source may appear, lasting for 
a few months, following the AIC event \citep{pk13}. 
In any case, these amounts are small enough to justify our assumption that the 
whole WD mass gets converted into the mass of the newborn NS (see Sect.~\ref{sec:MT}).\\ 
Another issue is that any NS formed via AIC may shortly afterwards begin to accrete additional material from its companion star, 
once this donor star re-fills its Roche~lobe when recovering from the dynamical effects of the implosion
(partly caused by the released gravitational binding energy in the transition from a WD to a more compact NS). 
Therefore, regardless of its initial properties, any NS formed via AIC could in principle subsequently be spun up to become an MSP,
as suggested by \citet{hrs83}.
The problem is, as pointed out by \citet{htw+10}, that this post-AIC accretion phase should then resemble
the conditions under which 
normal, old NSs are spun up to become MSPs via the conventional channel and, consequently,
one cannot easily distinguish the outcome of this formation path from the standard scenario.

In this work, we therefore concentrate on answering the second and the third questions raised above, 
i.e. if MSPs can be produced via AIC events which are immediately followed by subsequent mass transfer 
and, if so, how they can be distinguished observationally 
from those MSPs formed via the standard SN channel. 
We aim at investigating which progenitor binaries lead to AIC in the first place 
and we present detailed modelling of both pre- and post-AIC evolution to predict
the properties of MSPs formed via the indirect AIC channel.
The structure of our paper is as follows: 
In Sect.~\ref{sec:AICobs} we review the suggested observational evidence for NS production via AIC.
The computer code and our assumptions governing
the pre- and post-AIC mass-transfer processes are given in Sect.~\ref{sec:MT}. In Sects.~\ref{sec:AICMS},
~\ref{sec:AICgiant}, and ~\ref{sec:AICHe} we present those of our calculated systems which successfully lead to AIC 
with main-sequence star, giant star, and helium star
donors, respectively, and review the properties of the binary pulsars formed. 
We discuss our results in a broader context in Sect.~\ref{sec:discussion}
and summarize our conclusions in Sect.~\ref{sec:conclusions}.

\section{Observational evidence for AIC}\label{sec:AICobs}
The question of the origin of NSs is closely related to many of their observable parameters: spin, B-field, age, space velocity, 
and the nature of their companion star.
As already pointed out, the observational evidence suggested in the literature for NSs formed via AIC can be categorized into three groups.
We now review this evidence in more detail. 

\subsection{The role of NS kicks}\label{subsec:NSkicks}
It has been well established from observations of radio pulsar velocities that most NSs receive a momentum kick at birth \citep{ll94,hllk05}.
These kicks are possibly associated with SN explosion asymmetries and 
may arise from non-radial hydrodynamic instabilities (neutrino-driven
convection and the standing accretion-shock instability) in the
collapsing stellar core. These instabilities lead to large-scale anisotropies
of the innermost SN ejecta, which interact gravitationally with the proto-NS
and accelerate the nascent NS on a timescale of several seconds \citep[e.g.][]{jan12,wjm13}.
For the {\em entire} population of NSs, the range of required kick velocity magnitudes extends basically from 
a few $10\;{\rm km}\,{\rm s}^{-1}$ to more than $1000\;{\rm km}\,{\rm s}^{-1}$, in order to explain both the existence of NSs
residing inside globular clusters (which have small escape velocities, $v_{\rm esc}<50\;{\rm km}\,{\rm s}^{-1}$) as well as  
bow shocks and SN remnants associated with high-velocity pulsars.
However, when considering only {\em young} NSs in the Galactic disk 
the study by \citet{hllk05} is interesting; it revealed that the velocities of 
young ($<3\;{\rm Myr}$) radio pulsars are well described by a single Maxwellian distribution with a three-dimensional mean
speed of $\sim\!400\;{\rm km}\,{\rm s}^{-1}$. 
Furthermore, there are no detections of any low-velocity ($v_{\perp} < 60\;{\rm km s}^{-1}$) 
single radio pulsars with a characteristic age, $\tau < 1\;{\rm Myr}$.  
These facts indicate that NSs which formed recently in young stellar environments (the Galactic disk) received large kicks
and that iron core-collapse SNe of type II and type Ib/c therefore, in general, result in these large kicks.

On the other hand, about half of the approximately 300 known MSPs are detected in globular clusters \citep{rhs+05,frb+08}.
Obviously, pulsars retained in globular clusters (GCs) cannot have formed with large kicks since these clusters have small 
escape velocities, except in a few rare cases where an isolated low-velocity pulsar could form in 
a disrupted binary involving a large kick with a finetuned direction \citep[cf. Fig.~5 in][]{tt98}. 
It is therefore tempting to believe that many of these MSPs in GCs were not formed by iron core-collapse SNe.

\subsubsection{Electron capture SNe}\label{subsubsec:EC}
It seems clear that the lowest mass SN progenitors may not evolve all the way to form iron cores
\citep[see][for a recent review on pre-SN evolution of massive single and binary stars]{lan12}.
The final fate of these stars with ONeMg cores is an electron-capture SN (EC~SN), 
i.e. a collapse triggered by loss of pressure support owing to the
sudden capture of electrons by neon and/or magnesium~nuclei \citep[e.g.][]{nom84,wch98}.
Work by \citet{phlh08} shows that the initial mass range for EC~SNe is quite narrow, only about $0.25\;M_{\odot}$ wide, 
which would imply that some 4\% of all single-star SNe would be of this type. However, 
it has been suggested by \citet{plp+04} that EC~SNe could occur in close binaries for stars with masses
between $8-11\;M_{\odot}$ since these stars lose their envelopes via mass transfer before entering the AGB phase and thus avoid the
dredge-up and the consequent 
erosion of the CO core by this process. Therefore, these stars undergo EC~SNe rather than becoming ONeMg~WDs, the likely outcome of
most single stars of the same mass.
Furthermore, these authors argue that 
EC~SNe lead to prompt explosions (rather than slow, delayed neutrino-driven explosions)
that naturally produce NSs with low-velocity kicks \citep[see also][who proposed similar ideas]{vdh04}. The idea of 
different NS kick magnitudes comes from the discovery of two classes of Be/X-ray binaries
with significantly different orbital eccentricities \citep{prps02}. Furthermore, the low eccentricities and the low masses ($\sim\!1.25\;M_{\odot}$)
of second-born NSs in double NS systems supports this picture \citep[][and references therein]{spr10}.

\subsection{The role of young NSs in GCs}\label{subsec:youngGCs}
In Table~\ref{table:AICcandidates} we list a number of 
apparently young NSs (characterized by slow spin and relatively high B-fields) that are found in GCs.
The lifetime as an observable radio source is of the order of 100~Myr for a young (i.e. non-recycled) pulsar. 
Therefore, if these NSs had formed via iron core-collapse SNe their existence in GCs would not only be unlikely 
for kinematic reasons (as explained above), it would simply be impossible  
given that the stellar progenitor lifetimes of SNe~II and SNe~Ib/c are less than a few 10~Myr, much shorter than the
age of the many Gyr old stellar populations in GCs. 
Similarly, the nuclear evolution timescales of stars undergoing EC~SNe is of the order of $20-50\;{\rm Myr}$,
which is still short compared to the age of GCs, 
and for this reason also EC~SNe cannot explain the existence of young NSs in GCs today.
It is therefore clear that these NSs in GCs, if they are truly young\footnote{
For an alternative view, see Verbunt \& Freire (2013) who argue that these
NSs that appear to be young, are not necessarily young.}, 
are formed via a different channel.

\begin{table}
\center
\caption{Neutron stars that are candidates for being formed via AIC in a globular cluster (a--d) or in the Galactic disk (e--h), respectively.
         See text for explanations and discussion.} 
\begin{tabular}{lrcccc}
\hline 
\noalign{\smallskip} 
       {Object}   & {$P$} & {$B^{*}$}            & {$P_{\rm orb}$} & {$M_{\rm comp}^{**}$} & {Ref.}\\ 
           {}     & {ms}  & {G}                  & {days}          & {$M_{\odot}$}            & {}\\ 
\hline 
\noalign{\smallskip} 
 PSR~B1718$-$19   & 1004  & $4.0\times 10^{11}$  & 0.258           & $\sim\!0.10$             &  a \\
 PSR~J1745$-$20A  &  289  & $1.1\times 10^{11}$  & --              & --                       &  b \\
 PSR~J1820$-$30B  &  379  & $3.4\times 10^{10}$  & --              & --                       &  c \\
 PSR~J1823$-$3021C&  406  & $9.5\times 10^{10}$  & --              & --                       &  d \\
\noalign{\smallskip} 
\hline
\noalign{\smallskip} 
 GRO~J1744$-$28   & 467 & $1.0\times 10^{13}$  & 11.8            & $\sim\!0.08$             &  e \\
 PSR~J1744$-$3922 & 172 & $5.0\times 10^{9}$   & 0.191           & $\sim\!0.10$             &  f \\
 PSR~B1831$-$00   & 521 & $2.0\times 10^{10}$  & 1.81            & $\sim\!0.08$             &  g \\
 4U~1626$-$67     & 7680  & $3.0\times 10^{12}$  & 0.028         & $\sim\!0.02$             &  h \\
\noalign{\smallskip} 
\hline
\end{tabular}
\begin{flushleft}
  $^{*}$ B-field values calculated from Eq.(5) in \citet{tlk12} which includes a spin-down torque due to a plasma-filled magnetosphere. \\
  $^{**}$ Median masses calculated for $i=60^{\circ}$ and $M_{\rm NS}=1.35\,M_{\odot}$.\\
  a) \citet{lbhb93}; 
  b) \citet{lmd96}; 
  c) \citet{bbl+94}; 
  d) \citet{blt+11}. 
  e) \citet{phk+97}; 
  f) \citet{brr+07}; 
  g) \citet{sl00}; 
  h) \citet{ynv02}; 
\end{flushleft}
\label{table:AICcandidates}
\end{table}

\subsubsection{A strong link to AIC -- first piece of evidence}\label{subsubsec:AICGCs}
An AIC event is not very different from an EC~SN and it is therefore expected that also NSs
formed via AIC will receive a small kick (if any significant kick at all). 
For this reason formation via AIC
could explain the many NSs in GCs and, more importantly, also the young ages of some of these NSs.
As we shall demonstrate in this work, pre-AIC binaries may, in some cases, reach ages exceeding 10~Gyr before a low-mass giant companion star  
initiates Roche-lobe overflow (RLO) leading to the AIC event. Therefore, we would expect ongoing AIC events, and thus formation of newborn NSs,
in GCs even today.

\subsubsection{NS formation via the merger of two massive WDs?}\label{subsubsec:COWDmergers}
For the sake of completeness, we also mention that
the merger of two massive WDs may also produce a pulsar \citep{sn85,sn04}. 
This could be an alternative way of producing young pulsars in an old stellar population like a GC. 
Even binary pulsars may be produced this way in a GC, since in a dense environment a single produced NS can 
capture a companion star later on \citep{rhs+05}.
It is more difficult for this scenario to produce MSPs in binaries in the Galactic disk. This scenario 
would not only require an initial triple system origin, which is somewhat rare in the Galactic disk
(although recent work by \citet{rdl+13} suggests that at least 20\% of all close binaries have tertiary companions),
it would also require the remaining binary orbit to survive the dynamical effects of the merger event.
However, it should be noted that the merging double CO~WD event 
is also a key scenario (the so-called double-degenerate scenario) suggested as a progenitor of SNe~Ia \citep[cf.][]{web84,it84,ypr07,vcj10,pkt+12}.

\subsection{AIC candidates in the Galactic disk}\label{subsec:AICclassic}
The evidence for AIC is found not only in GCs. In Table~\ref{table:AICcandidates} we list
a number of Galactic disk binary NS systems which are postulated candidates for having formed via AIC. 
A common feature of these NSs is a slow spin and a relatively high B-field and 
an ultra-light ($\le 0.10\;M_{\odot}$) companion star in a close orbit.
The idea that the origin of some high B-field, slow spinning NSs 
(e.g. 4U~1626$-$67, Her~X$-$1, and PSR~B0820+08) is associated with AIC was originally suggested by \citet{tv86}. 
Although it was believed at that time that B-fields decay spontaneously on a timescale of only 50~Myr (and therefore
these NSs could not have much larger ages), many of these sources
remain good candidates for AIC today even though it has been demonstrated that pulsar B-fields can remain
high on much longer timescales \citep{kul86,bwhv92}. 
One reason why these NS systems remain good AIC candidates is the very small masses of their companion stars which indicate that 
a significant amount of material ($0.5-1.0\;M_{\odot}$) was transfered towards the compact object\footnote{A small
companion-star mass suggests that the previous (or ongoing) mass-transfer episode was (is) dynamically stable \citep{ts99,prp02}.
However, even observed radio pulsars which are thought to have evolved via a CE phase have $B<5\times 10^{9}\;{\rm G}$ \citep{tlk12},
in general much smaller that the B-fields of the NSs listed in Table~\ref{table:AICcandidates}.}. 
The paradox is therefore that these NSs still have high B-fields and slow spins
even though a significant mass transfer has occurred (see below).

\subsubsection{The role of accretion-induced B-field decay in NSs}\label{subsubsec:NS_Bfields}
There is solid observational evidence that the surface B-field strengths of NSs decrease with accretion
\citep{tv86,smsn89,vb94,vb95}. The exact mechanism for this process is still unknown. It may be related to
decay of crustal fields by ohmic dissipation and diffusion from heating via nuclear processing of accreted material \citep{rom90,gu94,kb97},
burial (screening) of the field \citep{zha98,czb01,pm07}, or decay of core fields due to flux tube expulsion from the 
superfluid interior induced by rotational slow-down in the initial phases of mass accretion \citep{sbmt90}; see also review by \citet{bha02}. 
Even a small amount of material accreted may lead to significant B-field decay, contradicting the observational
evidence that these binary NSs have accreted large amounts of material. 

\subsubsection{A strong link to AIC -- second piece of evidence}\label{subsubsec:AIC_Gal_NS}
A possible solution to the above-mentioned paradox of observing close binaries with ultra-light companions
orbiting high B-field NSs, 
would be if these NSs were formed recently via AIC during the very final stages of the mass-transfer process in a binary. 

The associated required finetuning of the AIC event to occur near the termination of the mass-transfer phase
is important for preventing accretion of significant amounts of matter 
after the formation of the NS (resulting in low B-fields and fast spin).
This finetuning problem will be investigated further in this paper. We note that the slow spins are expected from efficient loss 
of rotational energy due to the emission of magnetodipole waves from these high B-field NSs.

In Sect.~\ref{subsubsec:puzzling} we return to additional discussions of observational evidence for AIC in view of
our theoretical calculations, and also comment on a handful of recycled radio pulsars 
in the Galactic disk with puzzling characteristics.


\section{Numerical methods and physical assumptions of AIC}\label{sec:MT}
The numerical models presented in this work are divided into two parts:
1) evolution prior to AIC and 2) post-AIC LMXB evolution.
Both parts are computed with
a binary stellar evolution code originally developed by \citet{bra97} on
the basis of a single-star code \citep[][and references therein]{lan98}.
It is a one-dimensional implicit Lagrangian code which
solves the hydrodynamic form of the stellar structure and evolution equations \citep{kw90}. The evolution
of the donor star,  
the mass-transfer rate, and the orbital separation are computed simultaneously through an implicit coupling scheme 
\citep[see also][]{wl99} using the
Roche-approximation in the formulation of \citet{egg83}. To
compute the mass-transfer rate, we use the prescription of \citet{rit88}. In Sect.~\ref{subsec:giants} we discuss the limitations of this
description in wide-orbit LMXBs with giant donor stars.
We employ the radiative opacities of \cite{ir96}, which we interpolated in tables
as function of density, temperature, and chemical element mass fractions, including carbon and oxygen. 
For the electron conduction opacity, we follow \cite{hl69} in the non-relativistic case, 
and \cite{can70} in the relativistic case.
The stellar models are computed using extended nuclear networks
including the PP~I, II, and III chains and the four CNO-cycles.
In our default models we assumed a 
mixing-length parameter of $\alpha=l/H_{\rm p}=1.5$ \citep{lan91} 
and a core convective overshooting parameter of $\delta _{\rm ov}=0.10$.
We tested several models using $\alpha=l/H_{\rm p}=2.0$ which resulted
in only slightly larger final WD masses ($\sim\!1\%$) orbiting recycled pulsars 
in somewhat larger orbits (up to $\sim\!3\%$ increase in $P_{\rm orb}$).

If the accreting ONeMg~WD reached the limiting Chandrasekhar mass for a rigidly rotating WD
\citep[i.e. $M_{\rm Chan}=1.48\;M_{\odot}$, e.g.][]{yl05} we assumed it collapsed to form a NS. 
Differential rotation can persist if the timescale of angular momentum
transport is smaller than the accretion timescale in accreting WDs,
and leads to a critical mass that is significantly higher than the canonical
Chandrasekhar mass \citep[e.g.][]{yl04}. 
However, magnetic torques resulting from the Spruit-Tayler dynamo may enforce nearly rigid rotation in accreting WDs with 
the considered accretion rates in this study. Further research is needed to verify this. 

The WD collapse was modelled both with and without a momentum kick imparted to the newborn NS.
Although AIC is generally believed to result in no kick, or possibly a small kick, we 
applied three different kick values ($w=0, 50, 450\,{\rm km\,s^{-1}}$) to the newborn NS.
Given that the physics behind the kick mechanism is still unclear, we have included a few extra models
with a large kick of $450\;{\rm km}\,{\rm s}^{-1}$ to probe the extreme boundary conditions of our calculations. 
We solved for the combined effects of sudden mass loss and imparted momentum kick
on the orbital dynamics following \citet{hil83}. In all cases we assumed the mass equivalent of $0.20\,M_{\odot}$ was 
lost as released gravitational binding energy yielding a post-AIC NS gravitational mass of $1.28\,M_{\odot}$.
The post-AIC LMXB evolution was followed using the same computer code \citep[see also][for additional details]{tlk11,tlk12}.
For a more general discussion of MSP formation from LMXBs we refer to e.g. \citet{esa98,ts99,prp02,del08}.

\subsection{Accretion onto a white dwarf}\label{subsec:WDacc}
The mass-transfer process and the physics of accretion onto a WD has been described in a large number of papers, 
e.g. \citet{wi73,ns77,nom82,pk95,it96,lv97,hkn99,ldwh00,liv00,hp04,yl03,yl04,nskh07}, and more recently in \citet{hksn12,whe12,iss12,sti+12,nst13,dhbp13,mcc+13}. 
Most of these papers aim at investigating progenitors of type~Ia supernovae (SN~Ia) which are
important for cosmology studies of the accelerating Universe \citep[e.g.][]{rfc+98,pag+99}. In these progenitor
systems an accreting {CO}~WD reaches the Chandrasekhar limit and explodes. Observationally, these systems manifest themselves as cataclysmic variables \citep{hel01},
symbiotic systems \citep{ken86}, and supersoft {X}-ray sources \citep{vbnr92,kv97},  
depending on the companion star mass, its evolutionary status, and the mass-transfer rate. It should be noted that
the classification scheme overlaps somewhat on certain aspects.
The cataclysmic variables (CVs) can be further subdivided
into several classes: (classical) novae, recurrent novae, dwarf novae, polars, AM~CNn stars etc., again depending on the nature of the companion star,
the orbital separation, as well as the magnetic field strength of the accreting WD and accretion disk morphology. 

The classical novae have only been reported to erupt once whereas the recurrent novae usually erupt every one or two decades. 
The symbiotic variable star binaries have low-mass red giant donors which transfer mass at a fairly low rate
via (beginning atmospheric) RLO. They undergo
nova-like outbursts which last for a few decades before decaying back to their original luminosity.
The luminous persistent supersoft {X}-ray sources, on the other hand, display luminosities between $10^{36}-10^{39}\,{\rm erg\,s^{-1}}$ 
revealing high mass-transfer rates of the order of $10^{-7}\,M_{\odot}\,{\rm yr}^{-1}$.
These systems often have a more massive main-sequence donor star undergoing thermal timescale mass transfer.

It is generally believed that the accreting WDs in supersoft X-ray sources can grow in mass by 
accretion since the thermonuclear fusion of accreted hydrogen can be fairly
stable with these high accretion rates, in contrast to the case of nova systems
where nova explosions (caused by a violent ignition
in a thin shell of degenerate hydrogen) may erode the WDs \citep{wgtr86,pom+13}. The supersoft X-ray
sources are therefore believed to represent progenitors of SNe~Ia, because many
of these systems may produce a Chandrasekhar mass CO~WD. 
However, existing multicycle computations of hydrogen accretion onto massive WDs at a high rate are
still somewhat controversial. In a recent study \citet{iss12} showed that the accumulated helium is completely lost in strong helium flashes,
thereby making SN~Ia and AIC impossible, whereas another study by \citet{nst13} concluded that WDs continue to grow toward the Chandrasekhar limit. 

A number of numerical simulations show that when a CO~WD approaches the
Chandrasekhar limit with accretion rates of the order of $\dot{M}_{\rm WD}\simeq 10^{-7} -
10^{-6}\;M_{\odot}\,\rm{yr}^{-1}$, a thermonuclear-runaway caused by carbon
burning can occur at central densities of about 
$\rho_{\rm c}=2-5\times10^{9}\;\rm{g~cm}^{-3}$~\citep[e.g.][]{nty84,yl03,lht+06},
which may result in a SN~Ia explosion.  
In case an ONeMg~WD mass grows to the Chandrasekhar
limit, electron-captures onto $^{24}$Mg and $^{20}$Ne make the central density
increase to about $10^{10}~\rm{g~cm}^{-3}$ before oxygen ignites at the center
\citep{mnys80,mn87,nom87}.  The consequent oxygen deflagration with
this high density cannot lead to a thermonuclear explosion because of very
rapid electron-captures onto heavy elements produced by the oxygen burning
\citep{mnys80,tw92}. Therefore, a NS is the most likely outcome in the case of the collapse of an ONeMg~WD.

Although it is generally believed that accreting CO~WDs reaching the Chandrasekhar mass limit lead to a SNe~Ia, and that
accreting ONeMg~WDs reaching this limit undergo AIC and produce NSs, the outcome could in some cases be the opposite 
 -- see Sect.~\ref{subsec:exim} for further discussion.

\subsubsection{Dependence on mass-transfer rates}\label{subsubsec:MTrates}
\begin{figure}
\begin{center}
\mbox{\includegraphics[width=0.35\textwidth, angle=-90]{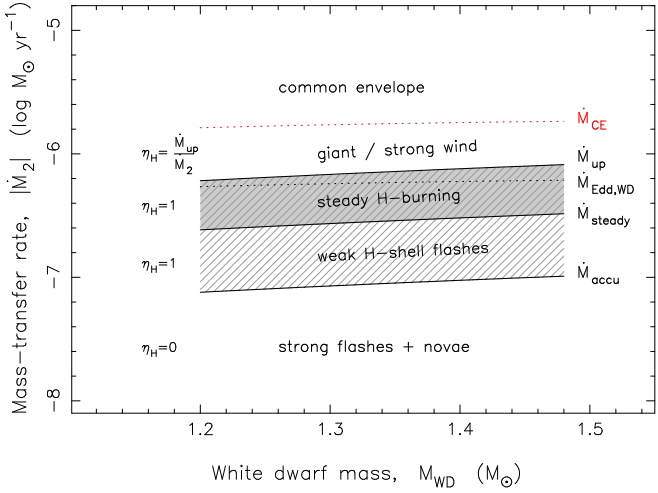}}
  \caption[]{
    The accretion window of the WD, the corresponding critical mass-transfer rates,  
    and our assumed mass-accumulation fractions, $\eta_{\rm H}$, here calculated 
    with a hydrogen abundance of $X=0.70$ of the accreted matter (see text).
    }
\label{fig:Mdot_crit}
\end{center}
\end{figure}
The response of the WD to mass transfer from the donor star depends on the mass-transfer rate, $|\dot{M}_2|$.
In this work we assume that the WD receives (but does not necessarily accrete) 
mass at the same rate as it is lost from the donor star, $|\dot{M}_2|$. 
White dwarfs which accrete faster than the rate at which their cores can grow ($\dot{M}_{\rm up}$) will puff~up their envelope to giant star dimensions (see below).
If the WD envelope does expand to such huge radii it will engulf the
donor star and the system is likely to evolve through a common envelope (CE) and spiral-in phase \citep{pac76,il93}.
The relevant critical mass-transfer rates for our work are (see Fig.~\ref{fig:Mdot_crit}):
   \[
      \begin{array}{lp{0.8\linewidth}}
         \dot{M}_{\rm CE}      & upper limit before the formation of a giant and a CE\\
         \dot{M}_{\rm up}      & upper limit for steady hydrogen burning\\
         \dot{M}_{\rm Edd, WD} & Eddington limit for spherical accretion\\
         \dot{M}_{\rm steady}  & lower limit for steady hydrogen burning\\
         \dot{M}_{\rm accu}    & lower limit for WD mass accumulation\\
      \end{array}
   \]
\noindent
where we have adopted the following {\em ad hoc} condition:
\begin{equation}
  \dot{M}_{\rm CE} = 3\, \dot{M}_{\rm Edd, WD} .
  \label{eq:MdotCE}
\end{equation}
The Eddington accretion limit 
is found by equating the outward radiation pressure 
to the gravitational force per unit area acting on the nucleons of the accreted plasma.
The radiation pressure (from photons that scatter on plasma electrons)
is generated from both nuclear burning at the WD surface and 
from the release of gravitational binding energy of the accreted material. 
The total energy production is given by: $L=(\epsilon_{\rm nuc}+\epsilon_{\rm acc})\,\dot{M}_{\rm WD}$,
where $\dot{M}_{\rm WD}$ is the accretion rate of the WD and $\epsilon_{\rm nuc}$ and $\epsilon_{\rm acc}$
denote the specific energy production from nuclear burning at the WD surface and release of gravitational binding energy, respectively. 
Hence, the Eddington accretion limit 
depends on the chemical composition of the accreted material and on the mass of the WD and is roughly given by:
\begin{equation}
  \dot{M}_{\rm Edd, WD} \approx  \left(5.5-6.2\right)\times 10^{-7}\;\,M_{\odot}\,{\rm yr}^{-1} 
  \label{MdotEddWD}
\end{equation}
for the WDs studied in this work. 

The value of $\dot{M}_{\rm CE}$ is probably the largest uncertainty in our modelling. 
The rationale of even considering an accretion rate $|\dot{M}_2|>\dot{M}_{\rm Edd, WD}$ 
is that the WD may drive a strong wind in a bipolar outflow which may prevent the WD envelope from otherwise expanding
into giant star dimensions \citep{nns79}. 
Many previous studies of accreting WDs have adapted the optically thick wind model 
of \citet{ki92} and \citet{kh94}, without any restrictions on $|\dot{M}_2|$ \citep[see][for a critique of this assumption]{cit98,ldwh00}. 
Given the uncertainties regarding the validity of this model, 
we adopt a maximum allowed mass-transfer rate limit of $|\dot{M}_2|= \dot{M}_{\rm CE} = 3\,\dot{M}_{\rm Edd, WD}$.
For most of our models we stopped the calculations if $|\dot{M}_2|>\dot{M}_{\rm CE}$ (assuming the system evolved into a CE and merged).
To test the dependence of this limit we also computed some models by allowing $\dot{M}_{\rm CE}=10\,\dot{M}_{\rm Edd, WD}$ for comparison.
As we shall see, the role of adapting the optically thick wind model or not 
has important consequences for the progenitor parameter space leading to AIC. 
Constraints of the wind mass loss from an accreting WD can be determined directly from radio observations of SN~Ia remnants \citep{csm+12}.
Hence, there is some hope that future observations can clarify the situation.

If a wind is driven from the WD (when $|\dot{M}_2|>\dot{M}_{\rm up}$) we calculate its mass-accumulation fraction 
using: $\eta _{\rm H} = \dot{M}_{\rm up}/|\dot{M}_2|$ 
to restrict the accretion rate to a maximum of $\dot{M}_{\rm up}$. This value represents the
upper limit for {\em steady} shell hydrogen burning of a WD and can be estimated from the growth rate of the degenerate
core in red giant stars undergoing hydrogen shell burning by applying the relation between 
core mass and luminosity \citep{it89}. As an example, \citet{hk01} found:   
\begin{equation}
  \dot{M}_{\rm up} = 5.3\times 10^{-7}\;\,M_{\odot}\,{\rm yr}^{-1}\; \cdot\left(\frac{M_{\rm WD}}{M_{\odot}}-0.40\right)
                     \left(\frac{1.7-X}{X}\right) , 
  \label{eq:Mdotup}
\end{equation}
which is valid for a hydrogen mass fraction $X\ge0.10$. There is a variety of similar expressions for this critical limit
in the literature \citep[e.g.][]{nom82,hkn96,hkn99,nskh07}. However, they do not differ by much and our results are stable
against these minor variations. (We note that $\dot{M}_{\rm up}\simeq \dot{M}_{\rm Edd, WD}$).\\
The minimum value for steady hydrogen shell burning is given by \citet{nom82}:
\begin{equation}
    \dot{M}_{\rm steady}=0.4\,\dot{M}_{\rm up} .
  \label{Mdot_steady}
\end{equation}
Although the hydrogen burning is not steady below this limit, the shell flashes are found to be weak
for accretion rates just slightly below the limit. 
Following \citet{hkn99} we therefore assume $\eta _{\rm H}=1$ in the entire interval:
$\dot{M}_{\rm accu} < |\dot{M}_2| < \dot{M}_{\rm up}$, where 
\begin{equation}
    \dot{M}_{\rm accu}=1/8\,\dot{M}_{\rm up} .
  \label{Mdot_accu}
\end{equation}
If the mass-accretion rate is below $\dot{M}_{\rm accu}$ violent shell flashes and nova outbursts cannot be avoided and thus 
the WD is prevented from increasing its mass (i.e. $\eta _{\rm H}=0$), or may even erode.

Following hydrogen burning the helium is processed into carbon and oxygen. The mass accumulation efficiency in helium
shell flashes was studied in detail by \citet{kh04}. We have adapted their mass accumulation efficiencies for helium burning, $\eta _{\rm He}$ into
our code.
The long-term effective mass-accretion rate of the WD is therefore given by:
\begin{equation}
  \dot{M}_{\rm WD} = \eta_{\rm H}\cdot \eta_{\rm He}\cdot |\dot{M}_2| .
  \label{MdotWD}
\end{equation}
For accretion of pure helium we used 
  $\dot{M}_{\rm WD} = \eta_{\rm He}\cdot |\dot{M}_2|$
and
\begin{equation}
  \dot{M}_{\rm up, He} = 7.2\times 10^{-6}\;\,M_{\odot}\,{\rm yr}^{-1}\; \cdot\left(\frac{M_{\rm CO}}{M_{\odot}}-0.6\right) ,
  \label{eq:Mdotup_He}
\end{equation}
where $M_{\rm CO}$ is the mass of the CO core of the helium donor star \citep{nom82}.
According to \citet{jhi93}, $\eta _{\rm He}$ might be somewhat smaller for direct accretion of helium, which leads to
stronger shell flashes, compared to the case where helium is accumulated via multiple cycles of hydrogen burning
(i.e. double shell burning). 

For recent discussions on the WD growth rate and the dependence on the WD mass and the mixing of the accreted material, see e.g. \citet{dhbp13} and \citet{nst13}.

\subsection{Orbital dynamics}\label{subsec:orbit}
We consider close interacting binary systems which consist of a non-degenerate (evolved) donor star and a compact object, 
in our case initially a massive WD and later on, in the case of an AIC event, a NS.
When the donor star fills its Roche~lobe, any exchange and loss of mass from such an {X}-ray binary will also 
lead to alterations of the orbital dynamics, via modifications in the orbital angular momentum,
and hence changes in the size of the critical Roche-lobe radius of the donor star. The stability of the
mass-transfer process therefore depends on how these two radii evolve (i.e. the radius of the star and its Roche-lobe radius).
The various possible modes of mass exchange and mass loss
include, for example, fast wind mass loss (Jeans mode), Roche-lobe overflow (with or without isotropic re-emission), 
and common envelope evolution \citep[e.g.][and references therein]{vdh94a,spv97}.
The RLO mass transfer can be initiated while the donor star is still on the main sequence (Case~A RLO), during hydrogen
shell burning (Case~B RLO), or during helium shell burning (Case~C RLO).
The corresponding evolutionary timescales for these different cases will in general proceed on
a nuclear, thermal, or dynamical timescale, respectively, or a combination thereof. This timescale is important for
the amount of mass that can be accreted and for
the extent to which the NS produced in the AIC can be recycled after its formation.  

The dynamical evolution of a binary system can be found by solving for the changes in the orbital separation, $a$.
The orbital angular momentum of a circular binary system is given by:
$J_{\rm orb} = \mu\,\Omega \,a^2$,
where $\mu$ is the reduced mass 
and the orbital angular velocity is:
$\Omega = \sqrt{GM/a^3}$.
A simple logarithmic differentiation of the orbital angular momentum equation yields the rate of change in orbital separation:
\begin{equation}
  \frac{\dot{a}}{a} = 2\,\frac{\dot{J}_{\rm orb}}{J_{\rm orb}} - 2\,\frac{\dot{M}_1}{M_1} 
                    - 2\,\frac{\dot{M}_2}{M_2} + \frac{\dot{M}_1 + \dot{M}_2}{M} ,
  \label{adot}
\end{equation}
where the two stellar masses are given by $M_1$ and $M_2$, the total mass is $M=M_1+M_2$, and the total change in orbital angular momentum per unit time is given by: 
$\dot{J}_{\rm orb}=\dot{J}_{\rm gwr}+\dot{J}_{\rm mb}+\dot{J}_{\rm ls}+\dot{J}_{\rm ml}$.
These four terms represent gravitational wave radiation, magnetic braking, other spin-orbit couplings, and mass loss, respectively
\citep[e.g.][and references therein]{tv06}.

In this work we adopt the so-called isotropic re-emission mode\footnote{Although at present there is not much observational evidence 
behind this model (not surprising given that systems with extremely high mass-transfer rates are short lived) there 
is some evidence of excess mass loss in the case of Cyg~X-2, as demonstrated by \citet{kr99} and \citet{pr00}.
This mass loss could have been be caused in the past by a relativistic jet (as observed in SS433) or by coronal winds from the outer parts of the accretion disk \citep{bb99}.}
for modelling the mass-transfer and the mass loss from the binary
\cite[e.g.][]{bv91,spv97}. We let the mass ratio between the donor star ($M_2$) and the accretor ($M_1$) be denoted by $q=M_2/M_1$.
Assuming that the direct wind mass loss of the donor star is negligible compared to the RLO mass-transfer rate
from the donor star, $|\dot{M}_2|$, and ignoring mass stored in a circumbinary torus, one can show that
a binary system always widens ($\dot{a}>0$) as a result of mass transfer if $q<1$. Similarly, a binary 
always decreases ($\dot{a}<0$) if $q>(1+\sqrt{17})/4\simeq 1.28$, 
irrespective of the amount of mass ejected from the vicinity of the accretor \citep[e.g. see Sect.~16.4.3 in][]{tv06}.
This fact is worth remembering when we analyse our results. 

In our calculations we have ignored any changes in $J_{\rm orb}$ due to magnetic braking or any other tidal spin-orbit interactions.
Magnetic braking is only known to operate efficiently in low-mass stars ($\la 1.5\;M_{\odot}$) which have convective envelopes.
As we shall see, our calculations show that only main-sequence donor stars with masses in the range $1.4-2.6\;M_{\odot}$ (depending on metallicity) 
transfer mass at the rates needed for AIC~events to take place. Therefore, to be consistent with all our calculations, we have not included magnetic braking
in the very few borderline cases. (These low-mass donors all have low-metallicities resulting in some stability against convective envelopes.)   
Our code includes gravitational wave radiation by calculating $\dot{J}_{\rm gwr}$ and correcting for it.
However, for these AIC progenitor systems the RLO mass-transfer timescales are always significantly shorter than the timescales on which gravitational wave 
radiation is important.

In all our calculations we assumed an initial (ONeMg)~WD mass of $M_{\rm WD}=1.20\;M_{\odot}$ prior to accretion.
As mentioned earlier, for these calculations the WD is simply treated as a point mass.

\subsubsection{The dynamical effect of the AIC on the orbit}\label{subsubsec:orbit_AIC}
If the WD mass reached $1.48\;M_{\odot}$ we assumed that the WD was subject to instantaneous AIC. 
The effect of sudden mass loss in a binary has been studied in detail by \citet{hil83} for bound systems, and
by \citet{tt98} for disrupted systems. Assuming a circular pre-AIC orbit we here follow \citet{hil83}  
to find the changes of the binary orbital parameters. The change of the binary semi-major axis, as a result of an asymmetric AIC, is given by:
\begin{equation}
  \frac{a}{a_0}=\left[ \frac{1-(\Delta M/M_0)}{1-2(\Delta M/M_0)-(w/v_c)^2-2\cos\theta\;(w/v_c)} \right] ,
  \label{eq:kick}
\end{equation}
where $a_0$ is the pre-AIC semi-major axis (radius), $a$ the post-AIC semi-major axis,
$\Delta M =0.20\;M_{\odot}$ the effective mass loss during the AIC when the $1.48\;M_{\odot}$
WD is compressed to a NS with a gravitational mass of $1.28\;M_{\odot}$ \citep{zn71}, $M_0=M_{\rm WD}+M_2$ the pre-AIC total mass,
$v_c=\sqrt{GM_0/a_0}$ the pre-AIC orbital velocity of the collapsing WD in a reference fixed on the companions star,
$w$ the magnitude of the kick velocity, and $\theta$ the angle between the kick velocity vector, $\vec{w}$ and the
pre-AIC orbital velocity vector, $\vec{v_c}$.
The post-AIC eccentricity is given by:
\begin{equation}
  e = \sqrt{1+\frac{2E_{\rm orb}J_{\rm orb}^2}{\mu\,G^2M_{\rm NS}^2M_2^2}} , 
  \label{eq:ecc}
\end{equation}
where the post-AIC orbital energy of the system is given by: $E_{\rm orb}=-GM_{\rm NS}M_2/2a$, and the orbital angular momentum is given by:
\begin{equation}
  J_{\rm orb} = a_0\,\mu\sqrt{(v_c+w\cos\theta)^2+(w\sin\theta\sin\phi)^2}   ,
  \label{eq:J_orb}
\end{equation}
where $\mu$ is the post-AIC reduced mass and $\phi$ is the angle between the projection of the
kick velocity vector onto a plane perpendicular to the pre-AIC velocity vector of the WD and the pre-AIC
orbital plane. We neglected any shell impact effects on the companion star since the amount of
material ejected in an AIC~event is expected to be negligible. 
Even for SN~Ib/c where a significant
shell is ejected, the impact effect on the orbital dynamics is small if the pre-SN
separation is larger than a few $R_{\odot}$ \citep{tt98}. 

Following the AIC we checked if the post-AIC periastron separation, $a(1-e)$ is smaller than the 
radius of the companion star, $R_2$. In that case we assumed that the system merges.
The post-AIC orbit is expected to be circularized with time and we therefore 
assumed that the tidal interactions reduced the semi-major axis by a factor
of $(1-e^2)$ in order to conserve $J_{\rm orb}$.
When no momentum kick was added ($w=0$) to the newborn NS, the relation between
post-AIC orbital separation (including the subsequent effect of tidal circularization), $a_{\rm circ}$
and the pre-AIC orbital radius, $a_0$ is simply given by \citep{vwv90,bv91}:
\begin{equation}
  a_{\rm circ} = a_0\frac{M_0}{M} = a_0\;\frac{1.48\;M_{\odot}+M_2}{1.28\;M_{\odot}+M_2} .
  \label{eq:a_circ}
\end{equation}

Since post-AIC evolution calculations also require a significant amount of computational time,
and each AIC~event can lead to a large set of possible parameter outcomes for $(w,\theta,\phi)$,
we have restricted ourselves to those cases which best probe the extreme cases for the post-AIC evolution
with respect to orbital periods and systemic recoil velocities resulting from the AIC~event. 

\begin{figure*}[t]
\begin{center}
  \includegraphics[width=0.60\textwidth, angle=-90]{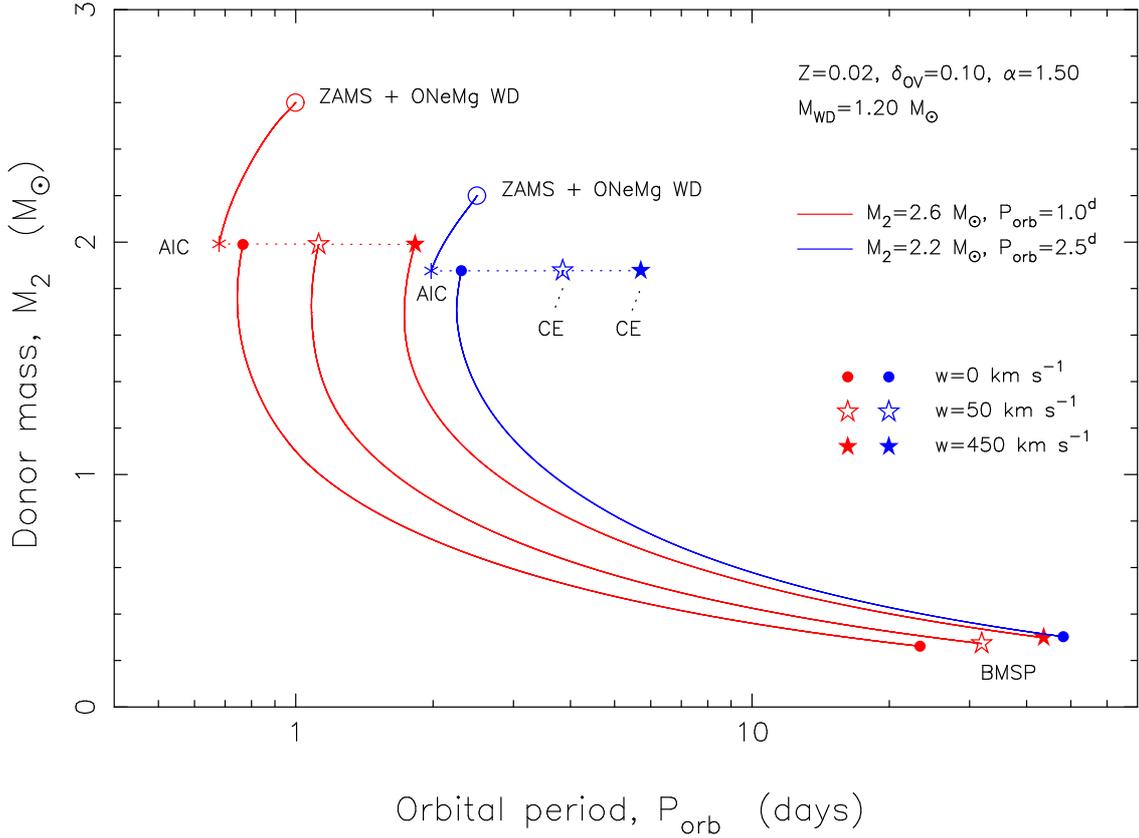}
  \caption[]{Evolutionary tracks in the ($P_{\rm orb},M_2$)-plane.
             The initial system configurations are: $M_2=2.6\,M_{\odot}$,
             $P_{\rm orb}=1.0\,{\rm days}$ (red tracks) and $M_2=2.2\,M_{\odot}$, 
             $P_{\rm orb}=2.5\,{\rm days}$ (blue tracks). The ONeMg~WD has
             an initial mass of $1.20\,M_{\odot}$ and grows to $1.48\,M_{\odot}$
             when the AIC occurs. After the AIC three tracks were computed for each system
             depending on the kick velocity applied in the AIC. Here we applied
             $w=0$, $w=50\;{\rm km s}^{-1}$ ($\phi=\theta=0^{\circ}$), and $w=450\;{\rm km s}^{-1}$ ($\phi=107^{\circ}$ and $\theta=90^{\circ}$)
             resulting in different values of $P_{\rm orb}$ for the post-AIC system.
             In two post-AIC cases presented here, and in general if the donor star is quite evolved by the time it refills its Roche~lobe
             (i.e. because the post-AIC $P_{\rm orb}$ is large) {\em and} if the donor star 
             is significantly more massive than the newborn NS ($1.28\,M_{\odot}$), the RLO  
             becomes dynamically unstable and the system evolves through a common envelope (marked by CE). 
    }
\label{fig:PM2-plane}
\end{center}
\end{figure*}
\subsection{Post-AIC LMXB evolution}\label{subsec:post_AIC_orbit}
The evolution of post-AIC binaries is, in principle, similar to normal LMXB evolution
(i.e. a donor star which transfers matter and angular momentum to an accreting NS). 
For this recycling process we follow \citet{ts99} and \citet{tlk12} for tracking
the evolution of the LMXB (see also Sect.~\ref{subsec:orbit} below).
The accretion rate onto the NS is assumed to be Eddington limited and is given by:
\begin{equation}
  \dot{M}_{\rm NS} = \left( |\dot{M}_2| -\max \left[|\dot{M}_2|-\dot{M}_{\rm Edd}\;,0\right] \right)\cdot e_{\rm acc}\cdot k_{\rm def} ,
  \label{MdotNS}
\end{equation}
where $e_{\rm acc}$ is the fraction of matter transfered to the NS which actually ends up being accreted and remains on the NS, 
and $k_{\rm def}$ is a factor that expresses the ratio of gravitational mass to rest mass of the accreted matter 
\citep[depending on the equation-of-state of supranuclear matter $k_{\rm def}\simeq 0.85-0.90$; e.g.][]{lp07}.
Here we assumed $e_{\rm acc}\cdot k_{\rm def}=0.30$.
Our motivation for this value is the increasing evidence of inefficient accretion in LMXBs, even in close systems where the
mass-transfer rate is expected to be sub-Eddington ($|\dot{M}_2| < \dot{M}_{\rm Edd}$) at all times
\citep[e.g.][]{jhb+05,avk+12}.
Possible mechanisms for inefficient accretion 
include propeller effects, accretion disc instabilities, and direct irradiation of the donor's atmosphere from the pulsar 
\citep{is75,vpa96,dlhc99}. 
For the post-AIC NS, we calculated the Eddington mass-accretion rate using:
\begin{equation}
  \dot{M}_{\rm Edd} = 2.3\times 10^{-8}\;\,M_{\odot}\,{\rm yr}^{-1}\; \cdot M_{\rm NS}^{-1/3}\cdot \frac{2}{1+X} .
  \label{MdotEdd}
\end{equation}

In Sect.~\ref{subsubsec:MSPspin} we specify a relation between the amount of mass accreted and the final pulsar spin period.

\subsection{Complete evolution in the ($P_{\rm orb},M_2$)-plane}\label{subsec:PM2-plane}
To demonstrate the orbital evolution: 1) during pre-AIC mass transfer, followed by 2) post-AIC mass transfer in an LMXB,
and to show the effect of possible kicks associated with the AIC, 
we have plotted complete evolutionary tracks in the ($P_{\rm orb},M_2$)-plane in Fig.~\ref{fig:PM2-plane}. 
The two donor stars in these examples ($M_2=2.6\;M_{\odot}$ and $M_2=2.2\;M_{\odot}$) are both more massive
than the accreting ONeMg~WD (initially with mass ratio, $q\sim\!2$.) This explains why both systems
decrease in $P_{\rm orb}$ prior to the AIC~event. At the moment of the AIC the orbits widen instantaneously as
a consequence of the sudden mass loss (c.f. Sect.~\ref{subsubsec:orbit_AIC}).
The larger the kick, $w$, the larger the post-AIC $P_{\rm orb}$ becomes at which the donor star refills its Roche~lobe and
continues mass transfer to the newborn NS in the LMXB source.
The final products are binary MSPs with He~WDs. During the LMXB phase the orbit changes from a converging system to
a diverging system when the mass ratio inverses (the exact value depends on the mass-transfer rate, c.f. Sect.~\ref{subsec:orbit}).

It is interesting that for the original $2.2\;M_{\odot}$ donor star,
the post-AIC mass transfer is not dynamically stable if the AIC was asymmetric (i.e. if $w=50\;{\rm km}\,{\rm s}^{-1}$
or $w=450\;{\rm km}\,{\rm s}^{-1}$).
In this case the post-AIC binary becomes wide enough that the donor star develops a deep convective envelope
before refilling its Roche~lobe. The result is that the subsequent mass-transfer stage leads to excessive mass-transfer rates and thus to the
formation of a CE. We did not follow the evolution of these systems further, although it is possible, in principle, that some
of the donor star envelopes would be loosely enough bound to allow for ejection during spiral-in
and thereby leave behind a mildly recycled pulsar orbiting a WD in a tight orbit.
We return to this possibility in Sect.~\ref{subsec:highBgal}.


\section{AIC in systems with main-sequence star donors}\label{sec:AICMS}
In the following three sections we present our results of AIC calculations in systems with main-sequence, giant, and helium star donors, respectively. 
We evolved a total of 240 binary systems. 
Key parameters from 36 examples of complete calculations (i.e. both pre-AIC and post-AIC LMXB computations leading to a recycled binary pulsar) 
are given in Table~\ref{table:AICmodels}.
Model names beginning with MS and MZ refer to main-sequence donor stars at solar metallicity ($Z=0.02$) and $Z=0.001$, respectively;
giant donor star models are denoted by GS (or GSZ for $Z=0.001$), and helium donor star models are denoted by He.
The given parameters for each model are the following: $M^{\rm ZAMS}_2$ and $P_{\rm orb}^{\rm ZAMS}$ refer to initial donor star mass and orbital period;
$t_{\rm RLO}$ is the age of the donor star when it initiates RLO; and $X_c$ is its central hydrogen content at that time;
$\Delta t_{\rm CV}$ is the duration of the mass-transfer phase until the AIC event; and $M_2^{\rm AIC}$ and $P_{\rm orb}^{\rm AIC}$ are
the donor star mass and the orbital period at the moment of the AIC. When a momentum kick ($w>0$) is added to the newborn NS, 
its magnitude and direction are given. The resulting systemic recoil velocity of each post-AIC system is given by $v_{\rm sys}$.
As a result of the AIC the system temporarily detaches. The time it takes until the donor star refills its Roche~lobe is denoted by $\Delta t_{\rm detach}$,
and $P_{\rm orb}^{\rm circ}$ is the orbital period at that time (after circularization). The duration of the subsequent post-AIC LMXB phase is given by
$\Delta t_{\rm LMXB}$. Finally, the parameters $M_{\rm WD}$, $P_{\rm orb}^{\rm MSP}$, $M_{\rm NS}$, $\Delta M_{\rm NS}$, $P_{\rm spin}$, and
$t_{\rm total}$ denote the WD mass, the orbital period, the NS mass, the amount of mass accreted by the NS during the post-AIC mass transfer, the final 
spin period of the recycled pulsar, and the total age of the binary system at this end point.

For hydrogen-rich donors we have initiated our calculations assuming a ZAMS star orbiting an ONeMg~WD. The error in placing the companion star on the ZAMS, 
and neglecting the evolution of this star while the ONeMg~WD forms, is not very significant. As we shall see in a moment, 
our main-sequence companions have maximum masses of $M_2=2.6\;M_{\odot}$, 
and even these stars evolve on a much longer timescale (at least by a factor of $\sim\!10$) compared to the typical $6-8\;M_{\odot}$ progenitor stars of ONeMg~WDs.

\begin{figure}
\begin{center}
  \includegraphics[width=0.78\textwidth, angle=-90]{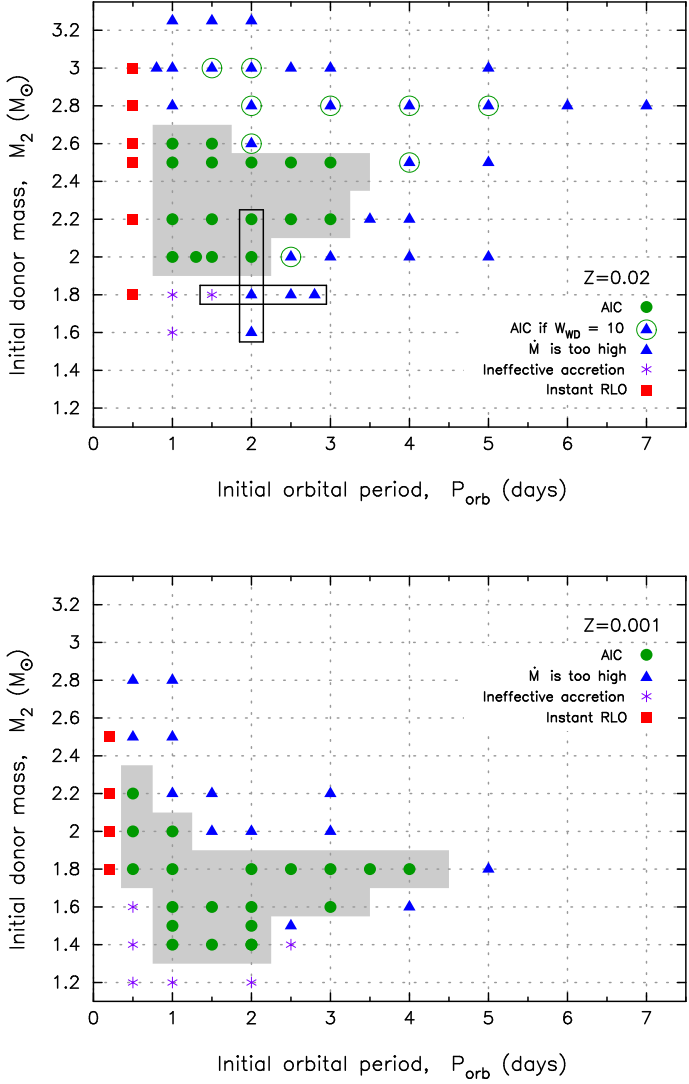}
  \caption[]{The grid of investigated initial orbital periods and masses for main-sequence donor stars with
             a metallicity of $Z=0.02$ (upper panel) and $Z=0.001$ (lower panel).
             The grey shaded region in each panel corresponds to systems which successfully evolve to the AIC stage (green circles).
             The blue triangles and purple asterisks correspond to cases where the mass-transfer rate was too high or too low,
             respectively, to allow for the WD to reach a critical mass of $1.48\;M_{\odot}$. The red squares indicate orbits which are too narrow
             to initially accommodate the ZAMS donor star. The blue triangles inside green circles in the upper panel are systems
             leading successfully to AIC assuming $\dot{M}_{\rm CE}=10\,\dot{M}_{\rm Edd, WD}$.
             Details of the mass-transfer process of the binaries inside the marked cross in the upper panel are represented in 
             Figs.~\ref{fig:Mdot_AICWDx} and \ref{fig:Mdot_AICWDy}.
    }
\label{fig:grid}
\end{center}
\end{figure}

\begin{figure}[h]
\begin{center}
  \includegraphics[width=0.35\textwidth, angle=-90]{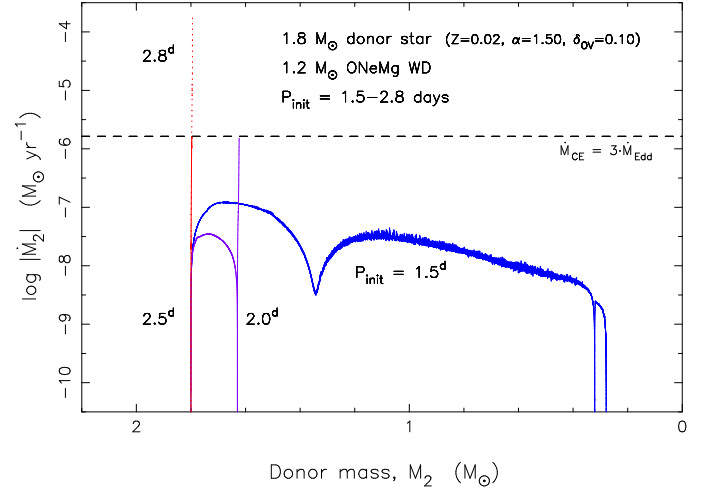}
  \caption[]{Mass-transfer rates of the four pre-AIC binaries shown in the horizontal part of the cross in
the upper panel of Fig.~\ref{fig:grid}.
The mass-transfer-rate from the
donor star in the system with $P_{\rm orb}=1.5^{\rm d}$ is too low (see Fig.~\ref{fig:Mdot_crit})
to cause the accreting WD to grow sufficiently in mass and trigger an AIC~event.
For the donors in the other systems, on the other hand, the mass-transfer rate is too high
to result in stable mass gain of the accretor. Hence, none of these models
resulted in a successful AIC. The differences in these mass-transfer rates can be understood
in terms of different thicknesses of convective envelopes (see text and Fig.~\ref{fig:kippenhahn}).
    }
\label{fig:Mdot_AICWDx}
\end{center}
\end{figure}

\begin{figure}[h]
\begin{center}
  \includegraphics[width=0.35\textwidth, angle=-90]{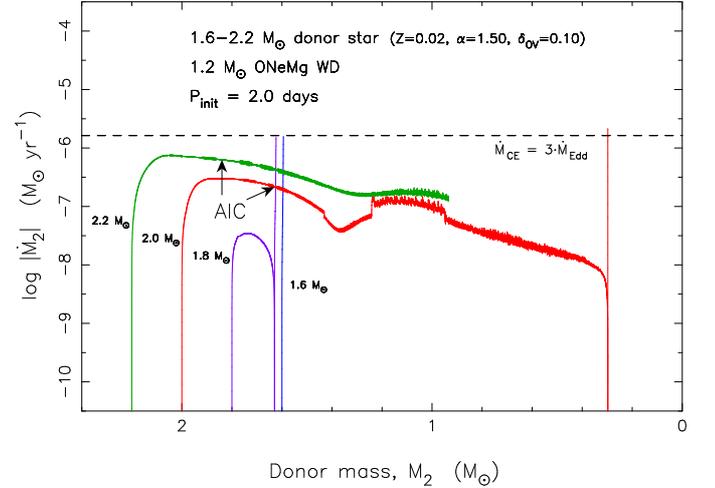}
  \caption[]{Mass-transfer rates of the pre-AIC binaries shown in the vertical part of the cross in
the upper panel of Fig.~\ref{fig:grid}. The four donor stars
have masses of $1.6-2.2\;M_{\odot}$ and $P_{\rm orb}=2.0^{\rm d}$.
The arrows mark the collapse of the accreting ONeMg~WD (AIC) for the two most massive stars.
Donor stars with initial masses $M_2\ge 2.6\;M_{\odot}$ result in excessive mass-transfer rates
and thus do not produce AIC~events (see text).
    }
\label{fig:Mdot_AICWDy}
\end{center}
\end{figure}

\subsection{Pre-AIC evolution with main-sequence donors}
In Fig.~\ref{fig:grid} we have plotted a grid of the initial orbital periods and donor star masses of
our investigated systems with main-sequence donor stars (i.e. supersoft {X}-ray sources).
The upper and lower panels are for different donor star metallicities. 
The type of symbol in each grid point represents the outcome of the computations, which we discuss in more detail below.
We note that the term main-sequence donor star is slightly misleading here since many of these stars have passed the termination age of
the main sequence (TAMS) by the time they fill their Roche lobes and become donors. Hence, many of these systems evolve via early Case~B RLO
from Hertzsprung-gap subgiant donors.
The systems which successfully evolve to the AIC event stage have initial orbital periods between $0.5-4\;{\rm days}$ and
initial donor star masses between $2.0-2.6\;M_{\odot}$ for a metallicity of $Z=0.02$ and between $1.4-2.2\;M_{\odot}$ for $Z=0.001$.

The shift in parameters in Fig.~\ref{fig:grid} is interesting, in particular in allowed donor star masses, 
which lead to AIC depending on the chemical composition of the donor star. The shift to lower donor star masses for lower metallicity 
can be understood from the smaller radii of these stars (due to their lower opacities) compared to stars with higher metallicity. 
Therefore, these stars become more evolved when eventually initiating their RLO, leading to higher values of $|\dot{M}_2|$ \citep[see also][]{ldwh00}.

\begin{figure*}
\centering
\includegraphics[width=\columnwidth]{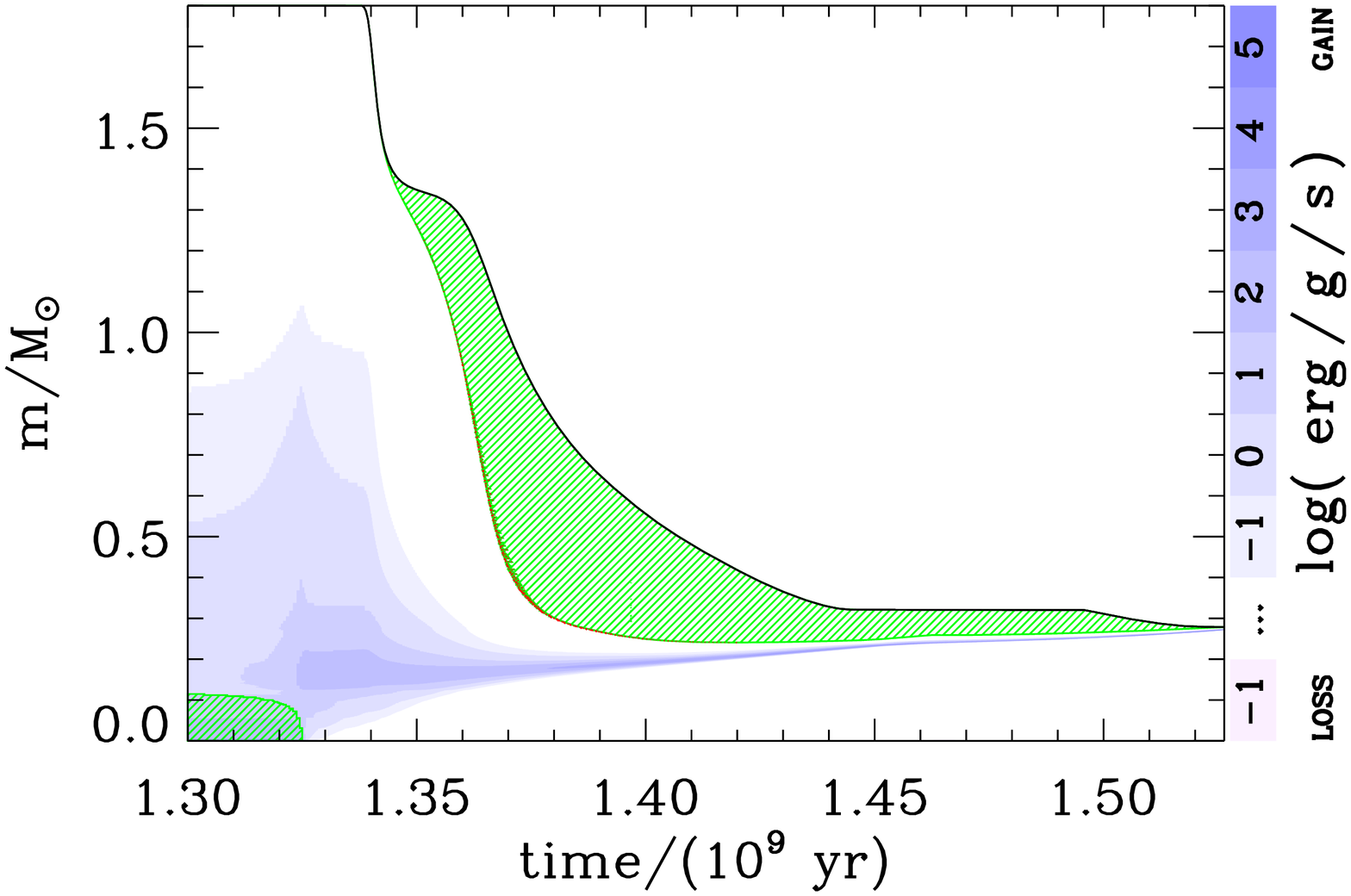}
\includegraphics[width=\columnwidth]{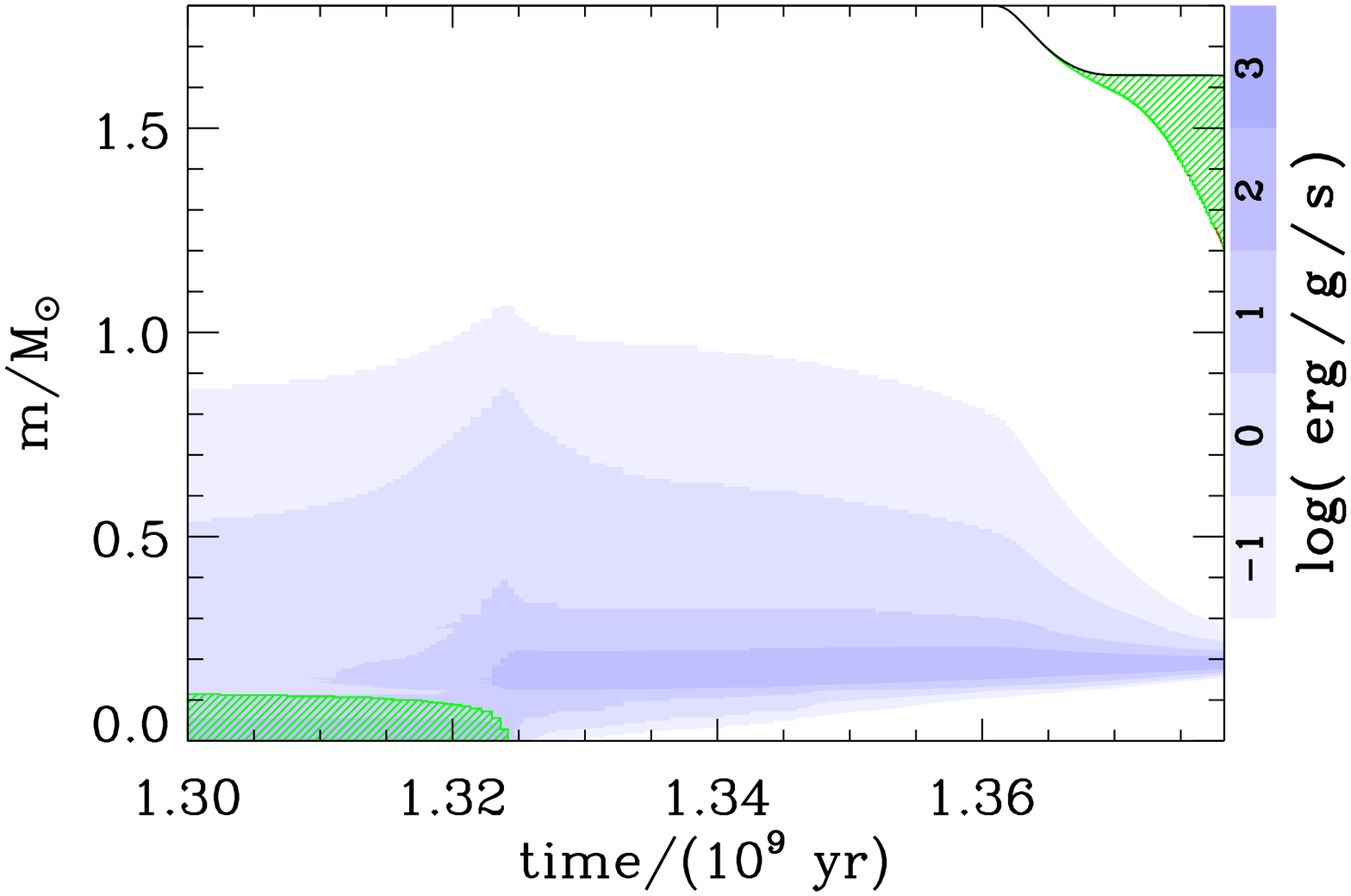}
\includegraphics[width=\columnwidth]{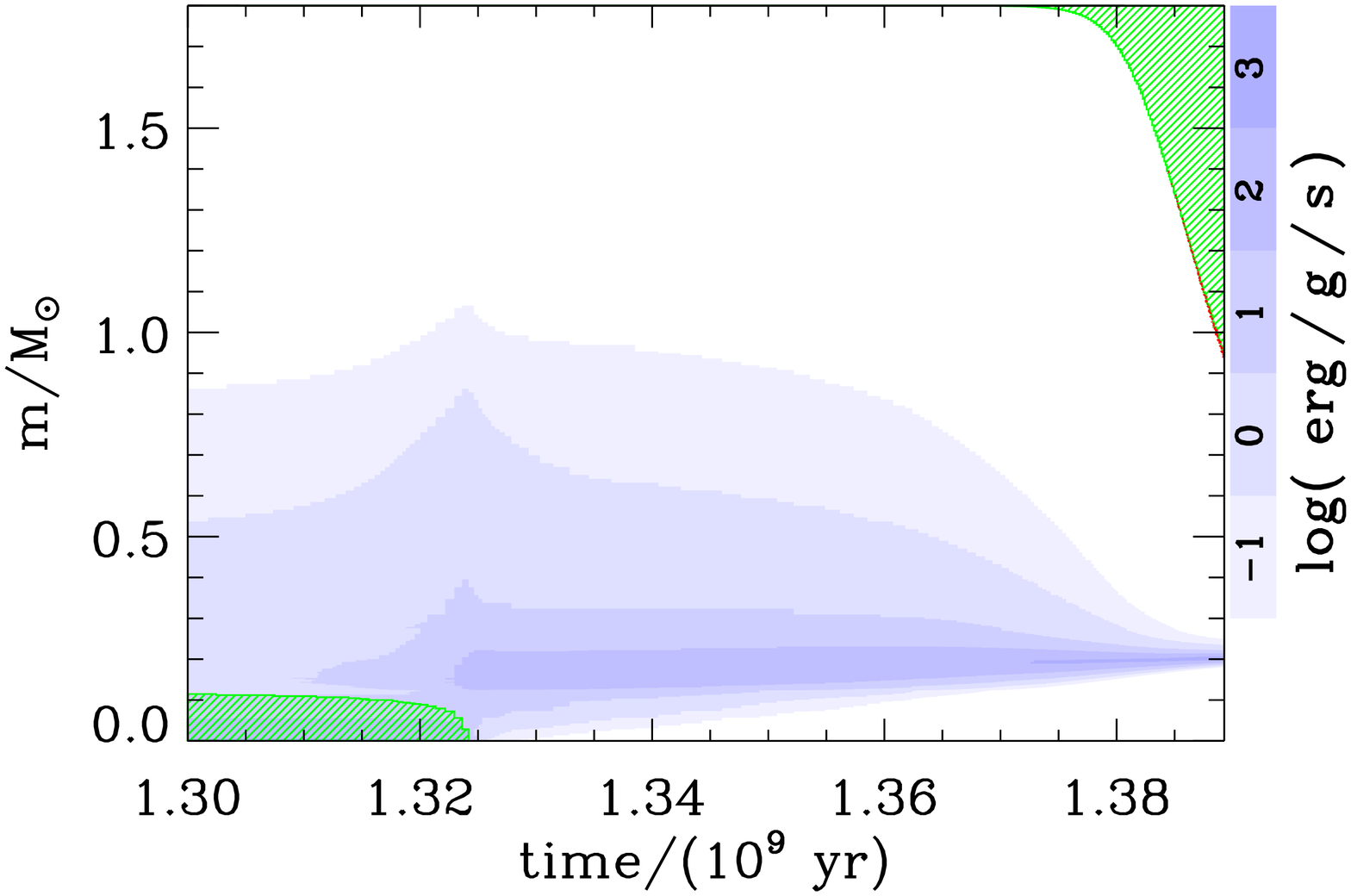}
\includegraphics[width=\columnwidth]{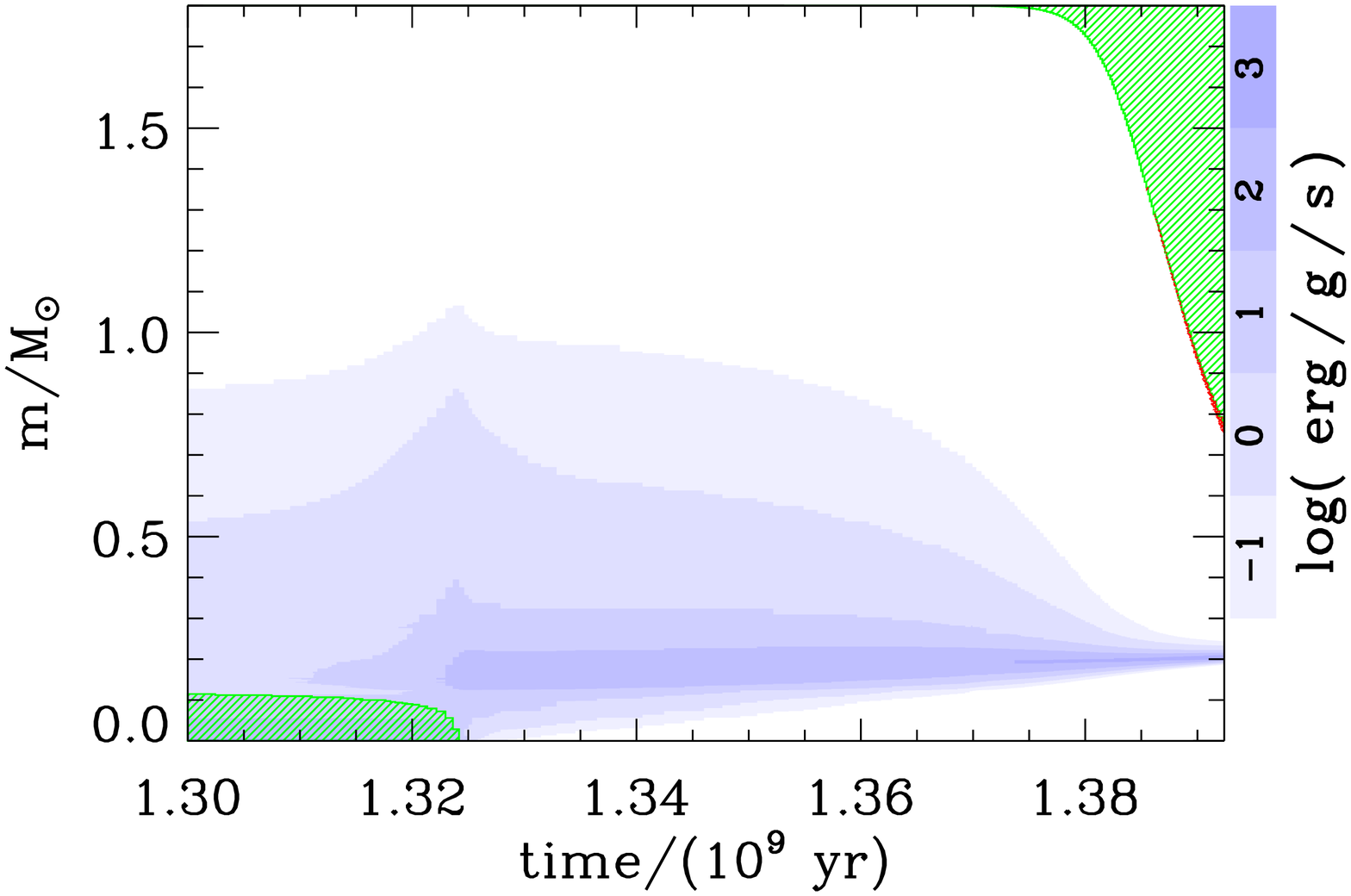}
\caption{
The Kippenhahn diagram of four $1.8\,M_{\odot}$ donor stars undergoing  
early Case~B RLO in {X}-ray binaries
with an accreting WD and orbital periods of 
1.5, 2.0, 2.5, and 2.8~days, respectively.
The plots show cross-sections of the stars in mass-coordinates
from the centre to the surface of the star, along the y-axis, as a function of stellar age on the x-axis.
The green hatched areas denote zones with convection (according to the Ledoux criterion) 
initially in the core and later in the envelope of the donor stars.
The intensity of the blue/purple color indicates the net energy-production rate; 
the hydrogen burning shell is clearly seen in all panels at $m/M_{\odot}\simeq 0.2$.
In the top-left panel, the donor star decreases its mass from $1.8\,M_{\odot}$
until it finally detaches from the Roche lobe and forms a $0.28\,M_{\odot}$ He~WD.
In the top-right panel, our calculation stops when the donor star
reached $1.62\,M_{\odot}$ since the mass-transfer rate became too high
when the donor star refilled its Roche lobe, following an initial phase of stable RLO with a lower value of $|\dot{M}_2|$.
In the two lower panels the mass-transfer rate went immediately up
to the critical value at the onset of the RLO.
The response of a donor star to mass loss depends
strongly on the depth of its convective envelope and thus, as seen here,
it depends on $P_{\rm orb}$. See Fig.~\ref{fig:Mdot_AICWDx} and text for a discussion. 
}
\label{fig:kippenhahn}
\end{figure*}

At first it may seem peculiar that neighbouring grid points can lead to a mass-transfer rate that is too low/too high
for the WD to grow sufficiently in mass (see e.g. $M_2=1.8\,M_{\odot}$,
$P_{\rm orb}=1.5^{\rm d}-2.8^{\rm d}$, and $Z=0.02$ along the horizontal part of the cross marked in the upper panel of Fig.~\ref{fig:grid}). 
However, this behaviour can be understood from the required finetuning of the WD accretion rate. 
In Fig.~\ref{fig:Mdot_AICWDx} we see that while the donor star in an orbit
with an initial $P_{\rm orb}=1.5^{\rm d}$ delivers an insufficient mass-transfer rate, the same system with 
$P_{\rm orb}=2.0^{\rm d}$ is seen to produce an excessive mass-transfer rate. The reason for this strong dependence on
$P_{\rm orb}$ is due to the corresponding rapid increase in the depth of the convective envelope with increasing radius of these 
shell hydrogen burning donor stars \citep{ps72}.
This is demonstrated in the Kippenhahn plot shown in Fig.~\ref{fig:kippenhahn}. The negative mass-radius exponents ($\zeta = \partial \ln R/ \partial \ln M$) 
of convective envelopes cause these stars to expand in response to mass loss and thereby result in excessive mass-transfer rates. 
Only in the first case for $P_{\rm orb}=1.5^{\rm d}$ (upper-left panel of Fig.~\ref{fig:kippenhahn}) will the mass transfer remain stable
since here the mass ratio is inverted (causing the binary to widen) by the time the envelope has developed a deep convection zone.
Similarly, in Fig.~\ref{fig:Mdot_AICWDy} we have shown the results of our mass-transfer calculations along four vertical, neighbouring points
in the cross marked in the upper panel of Fig.~\ref{fig:grid}. These points correspond to $1.6\le M_2/M_{\odot}\le2.2$, and in all
cases $P_{\rm orb}=2.0^{\rm d}$.
Again the explanation for the different outcomes is the differences 
in the depth of the convective envelopes.
The donor stars with $M_2\la 1.9\;M_{\odot}$ require a more advanced evolution to fill their Roche~lobes
and therefore they develop deep convective envelopes 
before, or during, the RLO which leads to $|\dot{M}_2|$ being too large. 
Hence, of those systems with $P_{\rm orb}=2.0^{\rm d}$ and $Z=0.02$, only those binaries with
initial donor star masses of $2.0\le M_2/M_{\odot} \le 2.5$ make it to the AIC~event.

In general, for all initial $P_{\rm orb}$, donor stars with $M_2\ge 2.6\;M_{\odot}$ result in $|\dot{M}_2|$ being too large 
(unless the criteria in Eq.(\ref{eq:MdotCE}) is relaxed to allow for $\dot{M}_{\rm CE} = 10\, \dot{M}_{\rm Edd, WD}$ 
in which case we get AIC solutions up to $M_2\approx 3.0\;M_{\odot}$). The reason is that 
in these systems the mass ratio, $q>2$, and therefore their orbits become significantly tighter with RLO, resulting in a large value of $|\dot{M}_2|$ 
(see Sect.~\ref{subsec:orbit}). 

\subsection{Post-AIC LMXB evolution with main-sequence donors}
\begin{figure*}
\begin{center}
  \includegraphics[width=0.40\textwidth, angle=-90]{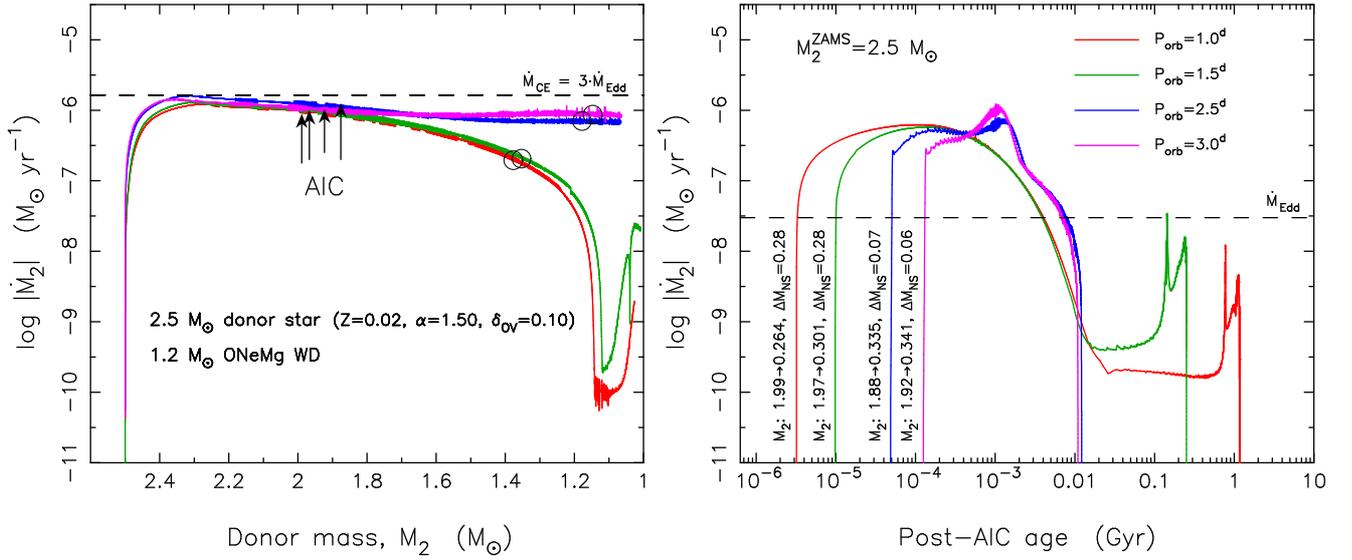}
  \caption[]{Left panel: Mass-transfer rates of four $2.5\,M_{\odot}$ donor stars 
             in pre-AIC binaries with initial orbital periods between $1.0-3.0\,{\rm days}$ leading to the post-AIC LMXB systems shown in the right panel. 
             The arrows mark the collapse of the accreting
             WD (AIC) and the evolution from this point onwards can be ignored since it is 
             now followed in the post-AIC LMXB systems shown in the right panel.
             The dashed line is the adopted upper limit for stable mass transfer (see Sect.~\ref{subsubsec:MTrates}).
             It is seen how the pre-AIC mass-transfer rates were quite close to our accepted upper limit.
             A slightly more massive donor star would lead to excessive mass-transfer rates
             and thus not result in an AIC~event (see also the systematic effect of an increasing donor mass in Fig.~\ref{fig:Mdot_AICWDy}).
             The open circles indicate hypothetical super-Chandrasekhar WD masses of $2.0\;M_{\odot}$.
             Right panel: Mass-transfer rates of the post-AIC LMXBs plotted as a function of time
             since the AIC~event.  
             The reason for the difference in time intervals between the AIC~event and the donor stars refilling
             their Roche~lobe (see initial interceptions of the four graphs with the x-axis) is mainly due to 
             different expansions of the orbit due to the AIC event, 
             caused by different $P_{\rm orb}$ at the moment of the WD collapse.
             The vertical text at the lower left side of each graph yields the
             mass of the donor star at the moment of the AIC, the mass of
             the final WD following the LMXB and the amount of mass
             accreted by the NS, $\Delta M_{\rm NS}$.
             The dashed line is the Eddington accretion limit for a NS.
             Only the systems with initial pre-AIC orbital periods of 1.0~days (red curve) and 1.5~days (green curve)
             experienced a long phase ($>100\;{\rm Myr}$) of post-AIC mass-transfer driven by hydrogen shell burning.
             This phase leads to significant accretion onto the NS
             (and effective recycling) since $|\dot{M}_2|< \dot{M}_{\rm Edd}$.
             Hence, the NSs in these systems were able to accrete more mass which results in 
             faster spinning MSPs (partly due to their smaller magnetospheres).
    }
\label{fig:Mdot_AICNS_12}
\end{center}
\end{figure*}
The post-AIC binary mass transfer resembles that of normal LMXB evolution with an accreting NS. The
only difference is that the donor star has already lost some of its mass during the pre-AIC evolution.
To model these LMXB systems we proceeded as explained in Sect.~\ref{subsec:post_AIC_orbit}. An example
of the results of this modelling is shown in the right panel of Fig.~\ref{fig:Mdot_AICNS_12} 
(the left panel shows the evolution of the pre-AIC binaries leading to these systems).
An interesting feature becomes clear when comparing the four evolutionary tracks.
Whereas the two donor stars in the widest orbits (initial $P_{\rm orb}=2.5^{\rm d}-3.0^{\rm d}$) are already 
undergoing shell hydrogen burning (Case~AB RLO) at the time of the AIC, the two donors in the 
shortest period systems (initial $P_{\rm orb}=1.0^{\rm d}-1.5^{\rm d}$)
are still undergoing Case~A RLO (core hydrogen burning) at the moment of the AIC. Hence, these latter systems remain LMXB (post-AIC) sources
on much longer timescales ($250\;{\rm Myr} - 1\;{\rm Gyr}$),  
initially via Case~A RLO, and later via Case~AB RLO once the hydrogen shell is ignited.  \\ 
As a consequence of the AIC event the binaries detach for $3\,000 - 100\,000\;{\rm yr}$ before the donor stars refill their Roche lobes
on a thermal timescale. (During pre-AIC mass loss the donor stars become smaller than their thermal equilibrium sizes.
After the AIC they expand to recover thermal equilibrium). 
This is a very short (thermal) time interval compared to the typical lifetime of a young pulsar (about $50-100\;{\rm Myr}$) and therefore
the possibility of detecting a system right at this epoch between the two long-lasting X-ray phases is quite small.
Indeed, none of the 6~known radio pulsars with a main-sequence companion are candidates for being post-AIC systems: in all
cases they have $P_{\rm orb}>50^{\rm d}$ and their companions are largely underfilling their Roche~lobes.

Next to each coloured graph in the right panel of Fig.~\ref{fig:Mdot_AICNS_12} is listed the mass of the donor star
at the moment of the AIC, $M_2^{\rm AIC}$, the final mass of the WD remnant orbiting the recycled pulsar, $M_{\rm WD}$, 
and the amount of mass accreted by the pulsar, $\Delta M_{\rm NS}$ (assuming an accretion efficiency of 30\% at sub-Eddington mass-transfer rates,
see Sect.~\ref{subsec:post_AIC_orbit}). The differences between the two main mass-transfer histories mentioned above 
(post-AIC Case~A RLO vs Case~AB RLO) is reflected in both 
$\Delta M_{\rm NS}$ and $M_{\rm WD}$. The consequences for the final MSPs systems will be discussed below.

\subsection{Resulting MSPs}
\subsubsection{Final orbital periods}\label{subsubsec:Pfinal}
\begin{figure}
\begin{center}
  \includegraphics[width=0.78\textwidth, angle=-90]{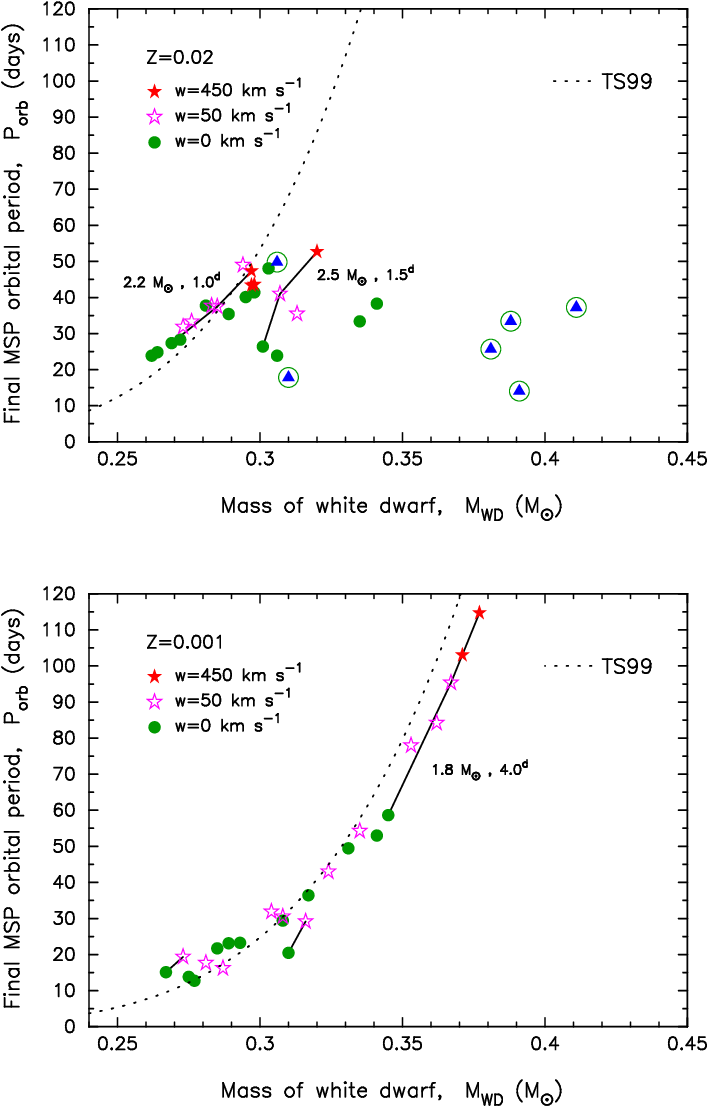}
  \caption[]{Final orbital period of the MSP binaries formed via AIC as a function of WD mass, evolving from 
             main-sequence donor stars
             with a metallicity of $Z=0.02$ (upper panel) and $Z=0.001$ (lower panel).
             The different symbols refer to different kicks during the AIC. The filled green circles correspond to $w=0$ (no kick),
             the open pink stars to $w=50\;{\rm km}\,{\rm s}^{-1}$, and the filled red stars are for $w=450\;{\rm km}\,{\rm s}^{-1}$.
             The symbols connected with a black line are for the same pre-AIC binaries but different values of the kick, $w$. 
             The dotted line is the ($M_{\rm WD},P_{\rm orb}$)--relation taken from \citet{ts99} and applies solely to low-mass
             donors with degenerate He~cores. The blue triangles in
             green circles are explained in Fig.~\ref{fig:grid}. See text for further details and discussions.
             }
\label{fig:MPrel_MS}
\end{center}
\end{figure}
In Fig.~\ref{fig:MPrel_MS} we have plotted our resulting binary MSPs in the final ($M_{\rm WD},\,P_{\rm orb}$)-plane.
The upper panel shows the resulting MSPs using a donor star metallicity of $Z=0.02$. The lower panel is for $Z=0.001$.
For clarity we have not included all systems shown in Fig.~\ref{fig:grid} which successfully evolved to the AIC, but we have included most systems and 
made sure to display those that yield the more extreme values of $M_{\rm WD}$ and $P_{\rm orb}$.
All the green filled circles were calculated assuming a symmetric AIC (i.e. $w=0$). The pink open stars represent AIC with a small kick magnitude of 
$50\;{\rm km}\,{\rm s}^{-1}$ and in all cases kick angles of $\theta=0^{\circ}$ and $\phi =0^{\circ}$. The red filled stars represent
systems surviving AIC with a large kick $w=450\;{\rm km}\,{\rm s}^{-1}$. In these cases the kick angles were always chosen to yield the widest 
possible post-AIC orbits from a systematic trial procedure. See Table~\ref{table:AICmodels} for examples. The black lines connecting three symbols
show the final results of three different kick values applied to the same AIC system.
One should keep in mind that AICs are most likely not accompanied with a kick (see Sect.~\ref{sec:intro}). However, we include the option here
for the sake of completeness.

From the lower panel of Fig.~\ref{fig:MPrel_MS} we note that 
for donor stars with low metallicity ($Z=0.001$) all of the final MSP systems fall approximately on
the well-known ($M_{\rm WD},P_{\rm orb}$)--relation \citep[e.g.][]{sav87,ts99}, shown as a dashed line. This relation follows from 
the relation between the radius of a giant star and the mass of its degenerate He-core \citep{rw71}. 
However, the more massive donors ($M_2>2.3\;M_{\odot}$) with non-degenerate cores do not obey this relation.
This explains why many of the systems in the upper panel ($Z=0.02$) deviate from the ($M_{\rm WD},P_{\rm orb}$)--relation.
Most of these donor stars leave behind relatively massive (hybrid) CO~WD remnants, see Sect.~\ref{subsubsec:BMSP-WD}.

An interesting outcome of these calculations is that the binary MSPs only form within a limited interval of $P_{\rm orb}$.
We conclude that the final orbital periods of MSPs, formed via AIC with main-sequence donor stars, 
are in the interval: $10^{\rm d}<P_{\rm orb}<60^{\rm d}$. Only in the unlikely case where large AIC kicks were applied
is it possible to form binary MSPs with $P_{\rm orb}$ up to 120~days.

\subsubsection{Nature of the final WD orbiting the MSP}\label{subsubsec:BMSP-WD}
\begin{figure*}
\begin{center}
  \includegraphics[width=0.45\textwidth, angle=-90]{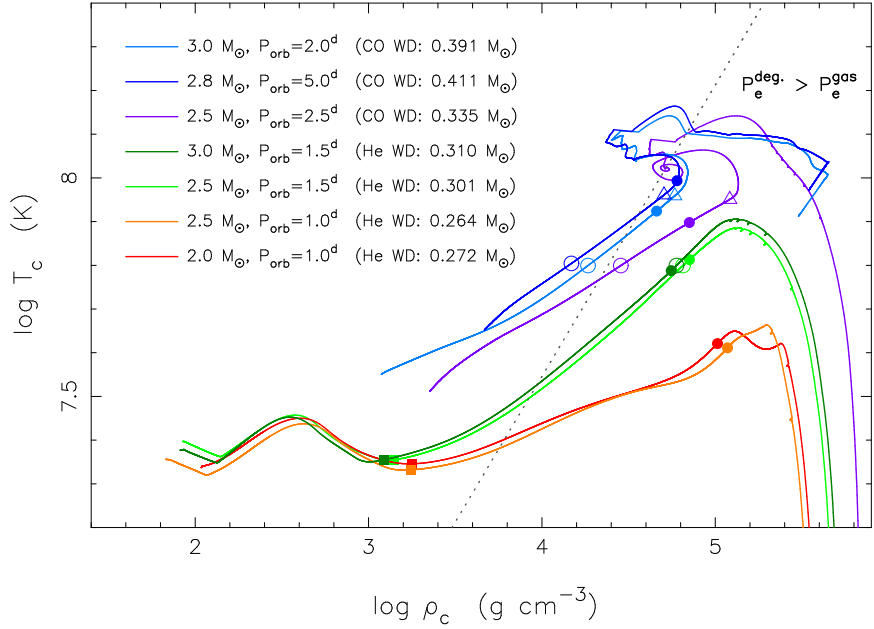}
  \caption[]{Evolutionary tracks in the core density--core temperature ($\rho_{\rm c},T_{\rm c}$)-plane.
             For each graph the initial (ZAMS) values of the donor star mass, $M_2$, the orbital period, $P_{\rm orb}$, 
             and in parenthesis the main chemical composition and the mass of the final WD are listed. The dotted line separates regions
             where the stellar pressure is dominated by the gas pressure (left) and the degenerate electron pressure (right).
             On each track the symbols represent: the termination of core hydrogen burning (filled squares), 
             the end of the RLO (filled circles), the onset of the $3\alpha$--process (open circles) and 
             the onset of efficient helium burning (i.e. when the luminosity generated by the $3\alpha$--process exceeds the energy loss rate by neutrinos, open triangles). 
             The calculations were followed to the WD cooling track, except for two
             cases causing numerical instabilities during hydrogen shell flashes.
             } 
\label{fig:rho-T}
\end{center}
\end{figure*}
To illustrate which donors leave He~WDs and which leave CO~WDs, we have plotted in Fig.~\ref{fig:rho-T} evolutionary tracks
in the ($\rho _{\rm c},T_{\rm c}$)-plane of WD progenitor stars with different masses and different values of $P_{\rm orb}$.
It is seen that the donors with lower masses and/or shorter initial orbital periods are more exposed to a high pressure of degenerate electrons and
a lower core temperature than
the more massive donors. Hence, these lighter donors leave He~WDs while the latter systems reach higher core temperatures and ignite helium to
produce CO~WDs. We find that the minimum threshold mass for efficient helium burning and production of a CO~WD is about $0.33\;M_{\odot}$, 
in agreement with previous work by e.g. \citet{kw90,tvs00,prp02}.
A more correct description of these low-mass CO~WDs in Fig.~\ref{fig:rho-T} of masses $0.335\;M_{\odot}$, $0.391\;M_{\odot}$, and $0.411\;M_{\odot}$ 
is hybrid WDs, since they have a composite composition of a CO core surrounded by a thick helium mantle \citep[e.g.][]{it85,ity97}.
In these stars, the temperature was never high enough to cause helium burning throughout the outer layers and thus the growth of the CO~core
was stalled when it reached a mass fraction of 50$-$75\%. Each of these WDs has a tiny hydrogen envelope of mass $1.2-3.7\times 10^{-4}\;M_{\odot}$
and evolved via one or more final hydrogen shell flashes due to unstable CNO burning in a thin hydrogen layer near their surfaces 
before they settled on the WD cooling track \citep[e.g.][]{asb01,ndm04}. As an example, our $0.335\;M_{\odot}$ hybrid CO-He~WD
underwent one last, vigorous hydrogen shell flash lasting only 15~yr, 
which caused it to fill its Roche~lobe and transfer $5\times 10^{-5}\;M_{\odot}$ towards the NS
with a maximum rate of $1.1\times 10^{-5}\;M_{\odot}\,{\rm yr}^{-1}$ (almost $\sim\!10^3\;\dot{M}_{\rm Edd}$).

\subsubsection{MSP spin periods}\label{subsubsec:MSPspin}
Given our calculated amounts of mass accreted by the NSs ($\Delta M_{\rm NS}$),  
we can constrain the possible pulsar spin periods after spin-up by using the formula \citep[see][for details]{tlk12}:
\begin{equation} 
\Delta M_{\rm NS} \approx 0.22\,M_{\odot}\; \frac{(M_{\rm NS}/M_{\odot})^{1/3}}{P_{\rm ms}^{4/3}}
\end{equation}
or 
\begin{equation} 
     P_{\rm ms} \approx 0.34\; (\Delta M_{\rm NS}/M_{\odot})^{-3/4} ,
  \label{eq:Pfit}
\end{equation}
where $P_{\rm ms}$ is the final equilibrium spin period in milliseconds, and $M_{\rm NS}$ is the 
initial NS mass (here, always $1.28\;M_{\odot}$ following the AIC). 
Obviously, it is not possible to spin~up a NS to faster rotation than its break-up spin period at $\sim\!0.6~{\rm ms}$, 
and our values do not include the circumstance of being limited by either gravitational wave radiation or a relatively large magnetospheric radius 
of the pulsar. Similarly, we discard the potential possibility of initially preventing post-AIC accretion onto the young energetic pulsar 
as a consequence of the so-called radio ejection mechanism \citep[i.e. ejection of material from the system caused by the outward magnetodipole radiation pressure
exceeding the inward ram pressure of material at the first Lagrangian point;][]{bpd+01}.
In Table~\ref{table:AICmodels} we list $\Delta M_{\rm NS}$ and $P_{\rm ms}$ for our calculated post-AIC LMXB systems.
It is seen that in almost all systems with main-sequence donor stars we obtain $\Delta M_{\rm NS} > 0.1\;M_{\odot}$ and therefore we find
that these MSPs will be fully recycled. \\
{\it We conclude that in all our models where MSPs formed via AIC, the predicted equilibrium spin periods of a few ms are identical to
those expected for MSPs formed via the conventional recycling channel where the NS was formed in a SN~Ib/c.} 

\subsubsection{MSP systemic space velocities}
We kept track of the post-AIC systemic velocities relative to the centre-of-mass rest frame of the pre-AIC binaries.
The systems receive a recoil due to the sudden mass loss, possibly combined with a smaller kick, during the AIC.
From conservation of momentum we obtain \citep[e.g. following][]{tb96}:
\begin{equation} 
     v_{\rm cm} = \sqrt{(\Delta P_x)^2 + (\Delta P_y)^2 + (\Delta P_z)^2} \;/ \;M ,
  \label{eq:recoil1}
\end{equation}
where the change in momentum is given by (cf. Sect.~\ref{subsubsec:orbit_AIC}):
\begin{equation}
   \begin{array}{lll}
   \vspace{0.2cm}
   \quad \Delta P_x   = M_{\rm NS}\,w\cos\theta - \Delta MM_2\sqrt{G/(rM_0)} \\
   \vspace{0.2cm}
   \quad \Delta P_y   = M_{\rm NS}\,w\sin\theta\cos\phi \\ 
   \quad \Delta P_z   = M_{\rm NS}\,w\sin\theta\sin\phi .\\
   \end{array}
\end{equation}
In Table~\ref{table:AICmodels} we also list the calculated systemic velocities of post-AIC binaries.
If large kicks (e.g. $w=450\;{\rm km}\,{\rm s}^{-1}$) were associated with AIC 
then the MSPs formed via this channel would reach velocities of up to 
$200\;{\rm km}\,{\rm s}^{-1}$, similar to the calculated systemic velocities of MSPs where the NS formed via a core collapse SN \citep{tb96}.
However, for $w=0-50\;{\rm km}\,{\rm s}^{-1}$ we find that the expected velocities of the resultant MPS formed via AIC
will be quite small and of the order of $10-30\;{\rm km}\,{\rm s}^{-1}$ at maximum\footnote{In Table~\ref{table:AICmodels} we always chose
$\theta = 0^{\circ}$ for applied kicks of $w=50\;{\rm km\,s}^{-1}$ which resulted in the largest possible post-SN values of $P_{\rm orb}$, but also in the
smallest values of $v_{\rm sys}$. For $\theta=180^{\circ}$ we obtain typical values of 
$v_{\rm sys}\approx 30\;{\rm km\,s}^{-1}$, for $w=50\;{\rm km\,s}^{-1}$.}.


\section{AIC in systems with giant star donors}\label{sec:AICgiant}
In Fig.~\ref{fig:grid_giant} we have plotted a grid of the initial orbital periods and donor star masses of
our investigated systems with giant star donors. These systems could potentially be detected as novae-like, symbiotic {X}-ray sources while 
the WD accretes material.
The giant star donors which successfully lead to AIC have masses between $0.9-1.1\;M_{\odot}$. The upper limit is set by the mass-transfer rate
(which is too high for larger values of $M_2$), and the lower limit is set by the age of our Universe ($0.9\;M_{\odot}$ donors
only evolve into giant stars within a Hubble-timescale of 13.7~Gyr if the metallicity is low enough).
The initial parameter~space dependence on metallicity is in general much weaker for these giant stars compared to the case of main-sequence donor stars
shown in Fig.~\ref{fig:grid}. The orbital periods, $P_{\rm orb}$ at the onset of the RLO must be at least $60-80\;{\rm days}$. If $P_{\rm orb}$
is smaller than this value the mass-transfer rate is too weak for the WD to grow efficiently in mass. 
The reason for different outcomes of $1.1\;M_{\odot}$ donor star models ($Z=0.02$) with $P_{\rm orb}=150^{\rm d}-190^{\rm d}$ is that 
$\dot{M}_2$ is very close to, and fluctuating near, our hard threshold limit, $\dot{M}_{\rm CE}$.

\subsection{Mass-transfer from giant star donors}
The mass-transfer modelling with giant star donors is difficult to calculate for two reasons. First, giant stars have low surface gravities
and thus extended atmospheres which make it non-trivial to estimate the turn-on of the RLO mass-transfer process. In some cases 
(see full discussion in Sect.~\ref{subsec:giants}) there is a
significant amount of mass transfer through the inner Lagrangian point while the donor star is still underfilling its Roche~lobe
(the so-called optically thin mass transfer).
Hence, for donor stars in the widest pre-AIC orbits ($P_{\rm orb}\ge 400^{\rm d}$) our code runs into problems, in particular for low-metallicity giant donors. 
Second, even low-mass giant stars have substantial wind mass loss \citep{rei75} which causes the orbits to widen prior to the RLO.
In some cases $P_{\rm orb}$ may increase by 20\% prior to the RLO while the donor star loses up to 30\% of its mass.
In our modelling, we have neglected the wind mass loss of the giant prior to RLO in order to isolate and better investigate the above-mentioned   
effect of optically thin mass transfer from these donor stars with extended atmospheres.

\begin{figure}
\begin{center}
  \includegraphics[width=0.78\textwidth, angle=-90]{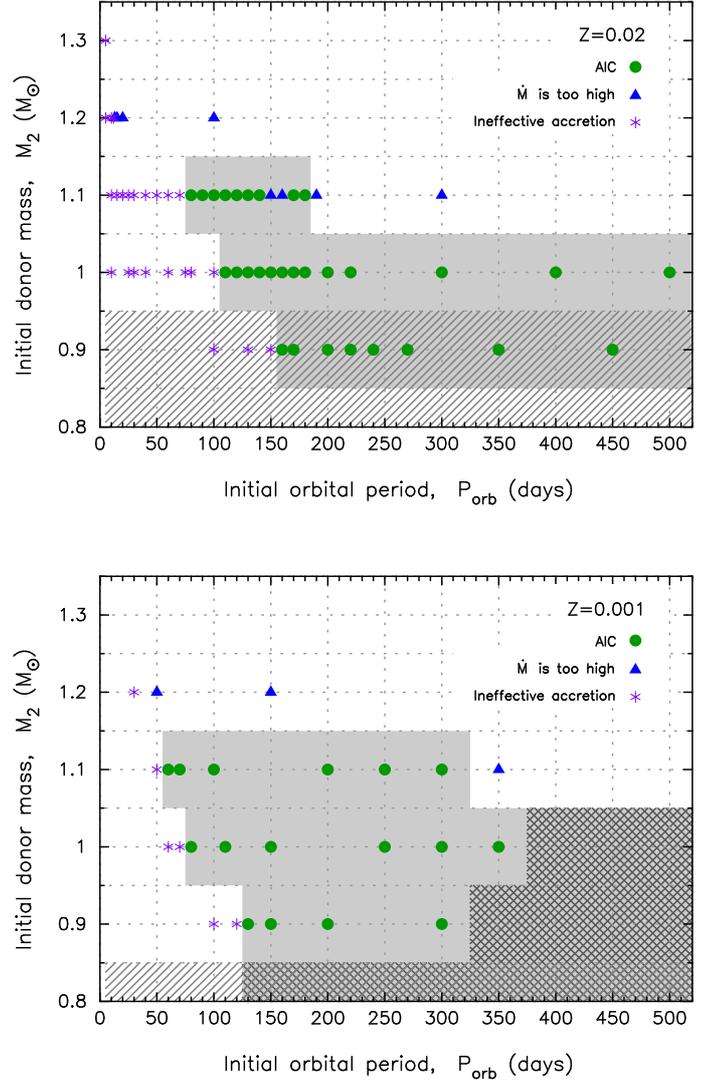}
  \caption[]{The grid of investigated initial orbital periods and masses for giant star donors with
             a metallicity of $Z=0.02$ (top) and $Z=0.001$ (bottom).
	     The meaning of the various symbols is equivalent to those in Fig.~\ref{fig:grid}. The grey shaded
             region in each panel corresponds to systems which successfully evolve to the AIC stage.
             Only giant donors with masses $0.9\le M_2/M_{\odot}\le 1.1$ are able to produce MSPs via the AIC channel.
             The Hatched regions indicate donor stars which evolve on a timescale longer than the Hubble-timescale.
             The cross-hatched region corresponds to giant donors which have so large atmospheric, hydrogen pressure scale-heights
             that our adopted mass-transfer scheme breaks down -- see text.
    }
\label{fig:grid_giant}
\end{center}
\end{figure}

As can be seen in Fig.~\ref{fig:grid_giant}  
for giant star donors, the transition from low values of $|\dot{M}_2|$ to donors which result in excessive  
values of $|\dot{M}_2|$ is very narrow for donor star masses, $M_2>1.1\;M_{\odot}$. Again the explanation is the sudden set-in
of rapid growth of the thickness of the convective envelope.
Including the effect of stellar winds is expected to shift the green points (successful AIC) in Fig.~\ref{fig:grid_giant} 
slightly to the left and upward. 

\subsection{Resulting MSPs}
The binary pulsars formed via AIC from giant star donors all end up in wide systems with $P_{\rm orb}>500^{\rm d}$.
The WD companions are either $0.40-0.46\;M_{\odot}$ He~WDs, or CO~WDs more massive than $0.46\;M_{\odot}$.
The pulsars are, in general, expected to be only mildly recycled with estimated 
spin periods of $10-500\;{\rm ms}$ due to the short mass-transfer phase, lasting $\Delta t_{\rm LMXB}\simeq 10^4-10^6\;{\rm yr}$,
following the AIC. Given this short phase of post-AIC mass transfer, combined with the effects of a young energetic pulsar 
(Sect.~\ref{subsubsec:MSPspin}), the final spin periods are expected to be somewhat slower than indicated above and in Table~\ref{table:AICmodels}. 
More notably, the systemic velocities of these binaries will be very small, $v_{\rm sys}\sim\!1\;{\rm km}\,{\rm s}^{-1}$,
given the general assumption that AICs are not accompanied by a momentum kick.
Hence, these will be at rest with respect to the local stellar population. \\
By the time the low-mass giant star donors initiate mass transfer towards the WD, they have already reached ages of the order
of 10~Gyr (see Table~\ref{table:AICmodels}). Therefore, we expect ongoing formation of NSs via the AIC channel in GCs today. 
In order for the giant star donors to deliver high enough mass-transfer rates to make the accreting WD grow sufficiently in mass,
the initial $P_{\rm orb}$ must be large. Hence, the orbital period at the moment of the AIC 
and the subsequent post-AIC LMXB orbit are also quite large. Such wide binaries enhance the likelihood of disruption by stellar encounters 
in a GC (e.g., Verbunt \& Freire~2013) and thus result in a number of isolated, young NSs in GCs as observed (or young NSs in close orbits
formed after a three body exchange event), see Table~\ref{table:AICcandidates}.\\
{\it We conclude that AIC is an attractive model to explain the small space velocities of some NSs and, in particular,
the retention of NSs in GCs. Furthermore, the AIC channel can explain the existence of relatively young NSs in GCs.}

\subsection{Break-down of the applied mass-transfer scheme for giant donors}\label{subsec:giants}
According to the canonical criterion for mass-transfer \citep{kw67} one simply has to ask whether the donor star radius, $R_2$ is 
larger or smaller than than its Roche-lobe radius, $R_{\rm L}$. Hence, one makes the implicit assumption that
the edge of the star is infinitely sharp and therefore that the mass transfer starts/ends abruptly rather than following a gradual transition. 
\citet{rit88} improved this criterion by taking the
finite scale height of the stellar atmosphere properly into account.
The mass loss from the donor star was modelled as a stationary isothermal, subsonic flow of gas which reaches
sound velocity near the nozzle at the first Lagrangian point, $L_1$. The accompanying mass-transfer rate is given by:
\begin{equation}
  |\dot{M}_2| = \;\rho _{\rm L1}\,v_{\rm s} \, Q \;
        \simeq \; \frac{1}{\sqrt{e}}\rho _{\rm ph}\,v_{\rm s}\,Q\,\exp\left(-\frac{\Delta R}{H_{\rm p}}\right) , 
  \label{eq:mtr}
\end{equation}
where $\rho _{\rm ph}$ is the gas density at the donors' photosphere, $v_{\rm s}=\sqrt{k T/(\mu m_{\rm H})}$
the isothermal sound speed, $Q$ the effective cross section of the flow at $L_1$ \citep[see ][]{mmh83}, 
and $\Delta R \equiv R_{\rm L}-R_2$. The last parameter,
\begin{equation}
  H_{\rm p} = \frac{kTR_2^2}{\mu m_{\rm H}GM_2}    ,
  \label{eq:scale-height}
\end{equation}
is the pressure scale-height of the stellar atmosphere ($\mu$ is the mean molecular weight).
This scheme was developed to study the turn-on (turn-off) of mass transfer in nearly semi-detached systems, the so-called
optically thin mass transfer for which $R_2 < R_{\rm L}$ \citep[see also][]{dmr89,kr90}\footnote{We note that $\Delta R$ and the mass ratio, $q$
are defined differently in \citet{rit88} and \citet{kr90}. The first paper has a typo in the last term in Eq.~(A8), which should be $f_2^{-3}(q)$.}. 
However, this mass-transfer algorithm was derived for low-mass main-sequence donor stars in CV binaries 
for which $H_{\rm p}\ll R_2$, and therefore mass transfer was only assumed to occur for $\Delta R \ll R_2$. \\
For giant stars this picture has to change \citep{pr89}.
These stars with low surface gravities often have $H_{\rm p}/R_2 \simeq 0.04$ and as a result we find that they can in some cases
cause mass-transfer rates above $10^{-7}\,M_{\odot}\,{\rm yr}^{-1}$ even for $\Delta R = 0.3\,R_2$ (i.e. while the donor
star is still underfilling its Roche lobe by 23\% in radius).  
Hence, for giant star donors the assumptions behind the original Ritter scheme breaks down. 
As a consequence, we did not allow for mass transfer with $\Delta R > 0.3\,R_2$ 
and our calculations were abandoned if this limit was reached
(see cross-hatched region in Fig.~\ref{fig:grid_giant}). 

As mentioned previously, we did not include wind mass loss in our models. This effect would cause the binaries to widen further
and thereby stabilize the systems against dynamical unstable mass transfer \citep{pr89}.
Another uncertainty is the effect of irradiation feedback on the long-term evolution of a compact binary \citep[e.g.][]{br04}.
However, the impact and the modelling of this effect, leading to cyclic accretion, is still unclear and is not included in our study.
Recent work by \citet{bdh12} on the evolution of ultra-compact X-ray binaries suggests that the inclusion of irradiation feedback
is not very significant for the final properties of these systems. Furthermore, for the wide-orbit LMXBs with giant donors \citet{rit08} 
argued that irradiation-driven mass-transfer cycles cannot occur since these systems are transient because of disc instabilities.


\section{AIC in systems with helium star donors}\label{sec:AICHe}
\subsection{Mass transfer from helium star donors}
\begin{figure}
\begin{center}
  \includegraphics[width=0.35\textwidth, angle=-90]{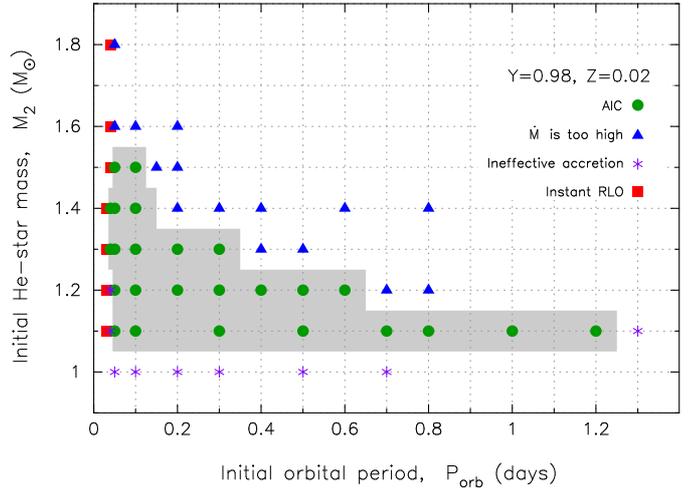}
  \caption[]{The grid of investigated initial orbital periods and masses for helium star donors with
             $Y=0.98$ and a metallicity of $Z=0.02$.
	     The meaning of the various symbols is explained in Fig~\ref{fig:grid}. The grey shaded
             region corresponds to systems that have successfully evolved to the AIC stage.
             See text for further discussions.
    }
\label{fig:grid_He}
\end{center}
\end{figure}

\begin{figure}
\begin{center}
  \includegraphics[width=0.35\textwidth, angle=-90]{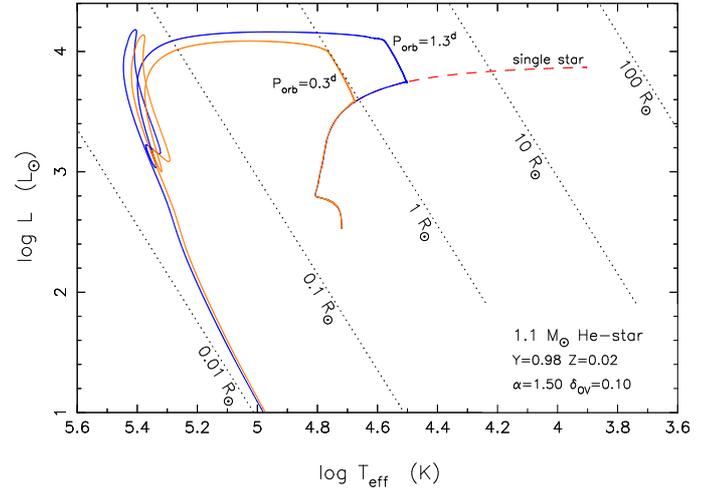}
  \caption[]{
             Evolutionary tracks in the HR--diagram for three $1.1\;M_{\odot}$ helium stars: two in binaries
             with a $1.2\;M_{\odot}$ WD and one evolving as a single star. The loops in the upper-left corner 
             of the two binary helium star tracks are caused by the ignition of shell helium burning. 
	     
    }
\label{fig:HeMT}
\end{center}
\end{figure}
Binary stars evolving from the ZAMS may lead to the formation of a tight binary with a massive WD, i.e. the remnant of the former primary star, 
and a helium star, i.e. the core of the secondary star that lost its envelope via CE evolution \citep[e.g.][]{lcwh10}. These systems will thereafter proceed
their post-CE evolution as described below, once the helium star fills its Roche-lobe (via so-called Case~BB RLO).

We constructed helium star donors with $Z=0.02$ (1.5\%~$^{14}$N, 0.2\%~$^{20}$Ne, and 0.3\%~isotopes of mainly Mg, Si, C, and O). 
As seen in Fig.~\ref{fig:grid_He}, the WDs that successfully accreted mass up to the AIC limit ($1.48\;M_{\odot}$)
had helium star donors with masses between $1.1-1.5\;M_{\odot}$ and initial $P_{\rm orb}$ between $1\,{\rm hr}\;(0.04\;{\rm days})$ and $1.2\;{\rm days}$, 
see Table~\ref{table:AICmodels} for details of 8~models. The mass-transfer rates become higher than the
critical limit ($\dot{M}_{\rm CE}$) for helium stars in relatively wide orbits.
However, for the lightest helium stars ($\le 1.1\;M_{\odot}$) this is not the case. Even in the widest orbits the mass-transfer rates become too low
to yield significant WD accretion because the orbits expand further in response to mass-transfer, given that $q<1$ at all times.

In the HR--diagram in Fig.~\ref{fig:HeMT} we compare the evolution of two $1.1\;M_{\odot}$ helium stars in binaries, with
initial $P_{\rm orb}=0.3^{\rm d}$ and $1.3^{\rm d}$, respectively, with the evolution of an isolated helium star of the same mass.
Whereas the single helium star evolves to a radius above $40\;R_{\odot}$ during shell helium burning, the binary helium stars which suffer from mass loss 
(and restriction in size due to their Roche~lobes) only evolve up to a radius of maximum $\sim\!2.6\;R_{\odot}$ (for an initial $P_{\rm orb}=1.3^{\rm d}$)   
before they evolve towards the WD cooling track as $0.80-0.82\;M_{\odot}$ CO~WDs.

We note that our models are computed with OPAL opacities which include carbon and oxygen abundances (cf. Sect.~\ref{sec:MT}).
Models computed with OPAL opacities \citep{irw92} that do not account for high C/O abundances (not shown here) 
resulted in final CO~WD masses that were 4--9\% larger than those in the models presented here.

\subsection{Resulting MSPs}
\begin{figure}
\begin{center}
  \includegraphics[width=0.35\textwidth, angle=-90]{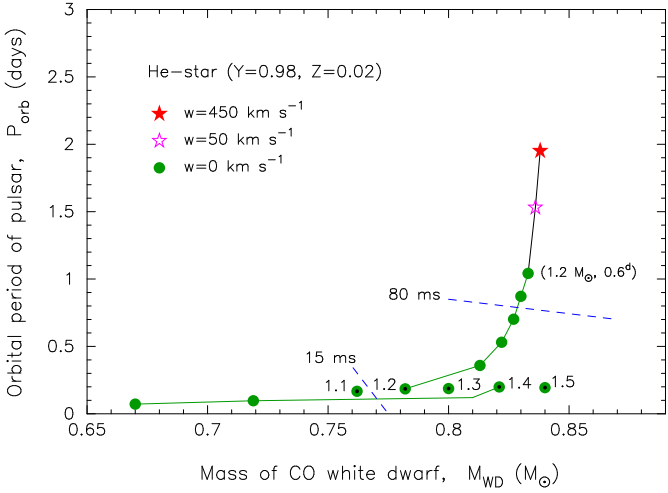}
  \caption[]{Final orbital periods of recycled radio pulsars as a function of their CO~WD companion masses for systems
             evolving from helium star donors 
             with a metallicity of $Z=0.02$. Data points connected with a green line were calculated from the same initial helium star mass
             ($1.2\;M_{\odot}$ and $1.4\;M_{\odot}$, respectively).
             Data points with black dots were all calculated from a binary with an initial orbital period of 0.1~days.
             The solid red star and the open purple star, connected with a black line to the green data points, 
             indicate similar calculations including an applied hypothetical kick velocity of $50\;{\rm km}\,{\rm s}^{-1}$ and
             $450\;{\rm km}\,{\rm s}^{-1}$, yielding the widest possible orbit.
	     The less massive the initial helium star donor, and the more narrow the orbit, the more material the NS accretes and the faster spin it obtains.
             The dashed lines indicate roughly where the pulsar is spun~up to 15~ms and 80~ms, respectively. See text for discussions.
    }
\label{fig:MPrel_AIC_He}
\end{center}
\end{figure}

It can be seen in Table~\ref{table:AICmodels} that the post-AIC LMXB phase, $\Delta t_{\rm LMXB}$ 
lasts for less than about 1~Myr (typically only $10^5\;{\rm yr}$), except for
one model (He1) where the RLO was initiated early on the helium star main sequence, resulting in 
$\Delta t_{\rm LMXB}=7.96\;{\rm Myr}$. Given the short lasting LMXB phase for these systems
the final pulsar masses are all close to their original post-AIC mass of $1.28\;M_{\odot}$. For the same reason, these pulsars
are only mildly recycled with minimum initial spin periods between $20-100\;{\rm ms}$ and we expect their  
B-fields to be larger than those of the fully recycled MSPs. (We note that the spin periods could be slower than stated in Table~\ref{table:AICmodels}
for the reasons given in Sect.~\ref{subsubsec:MSPspin}.) \\
The pulsar companions are all CO~WDs with masses between $0.6-0.9\;M_{\odot}$ and $P_{\rm orb}\simeq 2\;{\rm hr}-2\;{\rm days}$, 
see Fig.~\ref{fig:MPrel_AIC_He}. If a hypothetical large kick ($w>450\;{\rm km}\,{\rm s}^{-1}$) is applied the final
$P_{\rm orb}$ may reach 3~days. However, such large kicks are not expected for AIC (cf. Sect.~\ref{sec:AICobs}).

The tight binaries with initial helium star companions cause the post-AIC systemic velocities to reach about $20-30\;{\rm km}\,{\rm s}^{-1}$
(in models with no kicks) which is substantially faster than the recoil velocities imparted to the wider systems with
main-sequence or giant star companions, although much smaller than the velocities calculated for systems evolving
via the standard SN channel.

\begin{figure}
\begin{center}
  \includegraphics[width=0.35\textwidth, angle=-90]{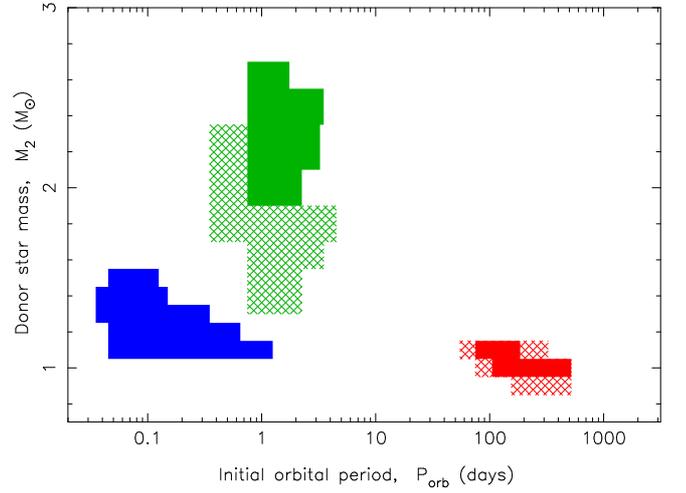}
  \caption[]{Grid showing all successful AIC systems, starting from an accreting $1.2\;M_{\odot}$ (ONeMg)~WD orbiting
             a main-sequence star (green), a giant star (red), or a helium star (blue) with a metallicity of $Z=0.02$. 
             Hatched areas are calculated for $Z=0.001$. The initial parameter space shown here is equivalent to that for
             progenitor binaries of SN~Ia with an accreting $1.2\;M_{\odot}$ (CO)~WD (see text). 
    }
\label{fig:grid_all}
\end{center}
\end{figure}

\section{Discussion}\label{sec:discussion}
Using a detailed stellar evolution code to probe the combined pre- and post-AIC evolution,
we have demonstrated that it is possible to form MSPs indirectly via the AIC channel. 
In Fig.~\ref{fig:grid_all} we show the initial parameter space of all binaries which successfully 
evolve to AIC according to our calculations.
In this section, we discuss our findings in more detail in relation to the predicted physical properties
of the resultant NS binaries and compare them with observations. We also discuss the connection between AIC and SNe~Ia progenitors and 
compare our work to other recent studies.

\subsection{Fully recycled MSPs or young NSs?}\label{subsec:outcome}
A general problem with postulating that a given observed high B-field NS was formed via AIC 
is that \citep[as pointed out by][and also demonstrated by \citet{sl00}]{wij97} 
it requires quite some finetuning to have the AIC occurring at the very final phases of the mass transfer in order to
explain the high B-field of the NS. If the WD collapses
earlier, the B-field of the newly formed NS (and its spin period, depending on $\dot{M}_2$) should decrease significantly
when the donor star refills its Roche~lobe following the AIC, thereby evolving through a relatively long lasting ($10^7-10^9\;{\rm yr}$) post-AIC LMXB phase.
In that case, all traces of its origin will be erased and the final NS cannot be distinguished from those recycled pulsars formed via the 
standard SN channel. Even accretion of a few $0.01\;M_{\odot}$ is enough to decrease the B-field significantly according to some models 
\citep[e.g.][and references therein]{wij97,zk06}.  
In all our model calculations the newborn NS undergoes post-AIC accretion, although in some extreme cases less than $10^{-3}\;M_{\odot}$ is accreted.

\begin{figure}
\begin{center}
  \includegraphics[width=0.35\textwidth, angle=-90]{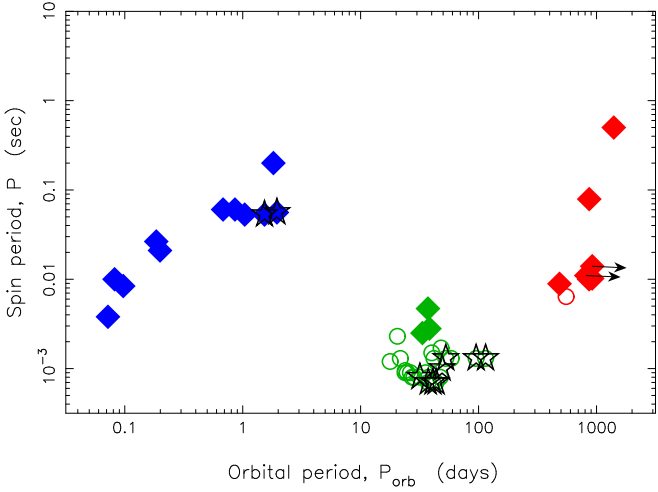}
  \caption[]{The Corbet~diagram for 55 MSPs (or mildly recycled pulsars) produced via AIC formation and subsequent accretion,
             including those listed in Table~\ref{table:AICmodels}.
             The three different donor star progenitor classes (green, red, and blue) are clearly distinguished,
             corresponding to main-sequence, giant star, and helium~star donors, respectively.
             Open circles are MSPs with He~WD companions, solid diamonds indicate MSPs with (hybrid) CO~WD companions.
             Symbols with a superimposed open black star correspond to AIC models where a kick was applied.
	     For a few systems with giant star donors only upper limits are given for $P_{\rm orb}$ (and $P$) 
             because of computational difficulties, cf. Sect.~\ref{subsec:giants}.
             }
\label{fig:AIC_corbet}
\end{center}
\end{figure}

\begin{figure}
\begin{center}
  \includegraphics[width=0.35\textwidth, angle=-90]{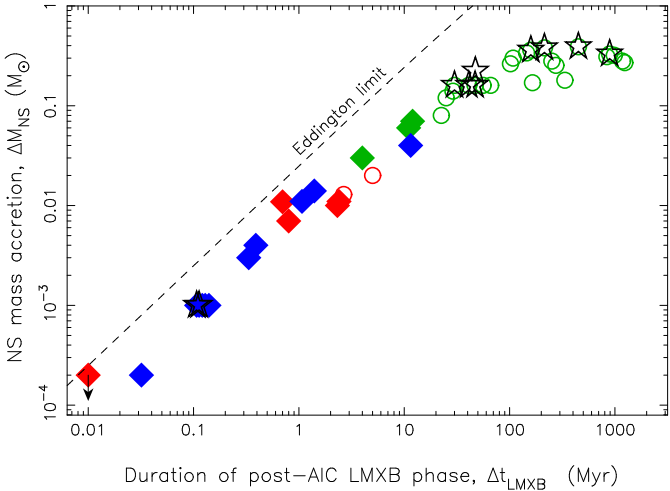}
  \caption[]{The amount of mass accreted by the NSs during the post-AIC LMXB phase, $\Delta M_{\rm NS}$ as a function
             of the duration of the post-AIC LMXB phase, $\Delta t_{\rm LMXB}$. 
             The symbols correspond to those in Fig.~\ref{fig:AIC_corbet}.
             Our calculations assumed an accretion efficiency of 30\%. The deviation from a straight line 
             is caused by the lower mass-transfer rates $|\dot{M}_2|\ll \dot{M}_{\rm Edd}$
             in systems with long mass-transfer timescales.
             The more mass a NS accretes, the faster is the final spin rate of the resultant MSP (see text).
             }
\label{fig:AIC_deltaM}
\end{center}
\end{figure}

\subsubsection{The AIC Corbet diagram}\label{subsubsec:AIC_corbet}
In Fig.~\ref{fig:AIC_corbet} we have plotted the Corbet diagram for a sample of  
our modelled NS systems which formed via the AIC channel (including the models listed in Table~\ref{table:AICmodels}). 
For systems with main-sequence donor stars, the resultant NSs eventually become fully recycled MSPs (i.e. with spin periods of a few ms and
most likely weak B-fields). Their orbital periods are confined to the interval: $10^{\rm d}<P_{\rm orb}<60^{\rm d}$
(up to 120~days when kicks are applied, cf. Sect.~\ref{subsubsec:Pfinal}). 
In Sect.~\ref{subsubsec:puzzling} we compare with the observed distribution of binary pulsars with He~WD companions.
For systems with either giant star or helium star donors, the spread in predicted final spin periods is much broader: from 4~ms to 0.5~sec.
However, the orbital periods are constrained to values of $P_{\rm orb}>500^{\rm d}$ and $P_{\rm orb}\le 2^{\rm d}$, respectively.

The predictions with respect to the values of the final spin periods are uncertain for systems with either giant star or helium star donors, 
since in both of these cases the post-AIC LMXB phase is short-lived.
On the one hand, if the NS in these systems only accretes $10^{-3}-10^{-2}\;M_{\odot}$ after its formation, 
we would expect it to form a mildly recycled pulsar with $P\simeq 4-500\;{\rm ms}$ as shown in Fig.~\ref{fig:AIC_corbet}.
On the other hand, in case the AIC produces (initially) very rapidly spinning and strong B-field pulsars,  
we cannot rule out that the radio ejection (Sect.~\ref{subsubsec:MSPspin}) or the propeller mechanism will prevent these NSs from
accreting matter during the short ($10^5-10^6\;{\rm yr}$) post-AIC LMXB phase. 
In that case, for these particular systems (red and blue symbols), the amount of material accreted by the post-AIC NS, $\Delta M_{\rm NS}$ could be 
substantially smaller (and the resulting spin periods longer) than modelled here 
(see Fig.~\ref{fig:AIC_deltaM} for our modelled values for $\Delta M_{\rm NS}$). 
Therefore, it might be possible to produce NSs via the AIC channel which retain relatively high B-fields and which therefore also appear young
(similar to what is observed, cf. Table~\ref{table:AICcandidates}).
To fully answer this question, one needs to model the
accretion disk--magnetosphere interactions in greater detail combined with a decisive model for the decay of the NS B-field. \\
Nevertheless, if the AIC formation channel is significant in terms of numbers
\citep[][conclude that the AIC route is just as important as the standard SN route]{htw+10}, 
then one may expect to observe a few cases where the AIC happened near the very end of the mass-transfer epoch, resulting
in a young, high B-field pulsar.

The orbital period distribution in Fig.~\ref{fig:AIC_corbet} is interesting. 
The resultant MSPs created via the AIC~channel preferentially form in certain orbital period intervals. 
The fully recycled MSPs mainly form with $10^{\rm d} < P_{\rm orb} < 60^{\rm d}$ (from systems with main-sequence
donor stars), whereas mildly recycled pulsars form in either very wide orbit systems with $P_{\rm orb} > 500^{\rm d}$ 
(from systems with giant star donors) or end up in close binaries with $P_{\rm orb} < 2.0^{\rm d}$
(all with massive CO~WD companions, left behind by helium stars donors). The tightest of these latter systems will merge within 1~Gyr.
Hence, according to our modelling there is a gap in orbital periods roughly between $60^{\rm d} < P_{\rm orb} < 500^{\rm d}$ 
where MSPs from AIC systems should not form 
(unless a kick is applied, in which case the lower limit increases to $100^{\rm d}$).
The reason why larger values of $P_{\rm orb}$ are not possible for systems with main-sequence star donors is that the mass-transfer rate becomes
too high for systems which could potentially widen to such large orbital periods. Only systems with low-mass giant star donors may
produce MSPs in wider systems. However, the final values of $P_{\rm orb}$ for these systems exceed 500~days. 
In Sect.~\ref{subsec:highBgal} we compare this modelled $P_{\rm orb}$ distribution with observed NS binaries.

\subsubsection{Production of young NSs in GCs}\label{subsubsec:AIC_GC}
In GCs, the possibilities are more favourable for producing young NSs which (initially) avoid recycling. The reason is that
some post-AIC binaries in GCs may become disrupted by an encounter event before the newborn NS experiences much accretion.
We note that for these systems the duration of the post-AIC detachment phase is comparable to the duration
of the subsequent LMXB phase (compare $\Delta t_{\rm detach}$ and $\Delta t_{\rm LMXB}$ in Table~\ref{table:AICmodels}).
Considering that GCs have an old stellar population, young NSs can only be produced from AIC with a giant star companion,  
since all possible main-sequence and helium star donors leading to AIC have nuclear evolution timescales 
much shorter than the age of the GCs (and thus these stars would already have evolved a long time ago).
Therefore, AIC events in GCs only occur in wide binaries which also enhances the probability of disruption before the post-AIC RLO 
has terminated, thereby strengthening the case for observing a young NS in a GC.\\
On a much longer timescale it is, of course, quite likely that these isolated NSs may capture a companion star and become recycled.
The LMXB IGR~J17480$-$2446 in Terzan~5 might be an example of a NS formed via AIC that has captured a new companion star and is now
undergoing recycling, as suggested by \citet{pavv12}.

\subsection{AIC vs SN~Ia: implosion or explosion}\label{subsec:exim}
All calculations in this work assumed an initial binary configuration with an accreting $1.2\;M_{\odot}$ WD, treated as a point mass. 
We implicitly assumed that when the (ONeMg)~WD reaches the Chandrasekhar mass at $1.48\;M_{\odot}$ \citep[the maximum mass of a rigidly rotating WD; e.g.][]{yl05} 
it collapses and forms a NS. However, predicting the final fate of accreting massive WDs is not trivial \citep[e.g.][]{nk91}. 
Accreting ONeMg~WDs reaching the Chandrasekhar limit may not always lead to an implosion that leaves behind a NS. 
The outcome depends on whether or not the effects of electron captures dominate over nuclear burning (oxygen deflagration) and 
hence it is sensitive to the central density where explosive nuclear burning is ignited \citep{nmsy79}. 
If the central density is too low, the timescale of electron captures is too long compared to that of the nuclear energy release
and the result may be explosive oxygen ignition and a SN~Ia.\\
Similarly, it may not always be the case that accreting CO~WDs explode in SNe~Ia when reaching the Chandrasekhar limit. 
This requires that the interior temperature (affected by heat inflow from surface layers) is high enough to ignite carbon
at a relatively low density. Therefore, the outcome depends on the duration of accretion phase and thus 
on the initial mass of the CO~WD. If the initial CO~WD mass is relatively high ($\ge 1.2\;M_{\odot}$), then the 
accretion phase leading to the Chandrasekhar limit is short and thus it may not result in a temperature high enough 
for such a (high-density) WD to explode in a SN~Ia \citep{nk91}.\\
Nonetheless, using our stellar evolution code we find that WDs formed with an initial mass of $1.2\;M_{\odot}$ in close binaries
are more likely to be ONeMg~WDs (or at least hybrid WDs with ONeMg cores embedded in a thick CO mantle)
compared to CO~WDs.

\subsubsection{SN~Ia progenitor calculations and shell impact}\label{subsubsec:SNIa}
Assuming instead that our point mass WD is a CO~WD leading to a SN~Ia (despite the uncertainties mentioned above), 
we have probed the progenitor parameter space for the single degenerate scenario with the results presented in this work
(see Fig.~\ref{fig:grid_all} and the discussion in Sect.~\ref{subsec:others}).
In addition, we have a library of many donor star structures, 
of different nature and at various evolutionary epochs, at the moment of the SN~Ia events  
(cf.~Table~\ref{table:AICmodels} for examples of stars for which we have detailed structures). 
These models can be used in future work to study the impact of the SN~Ia explosion on the donor star and thus help predict 
the expected properties of former companions stars when searching for these in SN~Ia remnants \citep[see e.g.][]{mbf00,dk12,lpr+12,prt10,prt12,prt13}.
These authors demonstrated that whereas helium star donors may only lose about 5\% of their mass, and main-sequence star donors lose some 10--20\%,
giant star donors may lose most of their envelope because of the SN impact. Since much less material is ejected in an AIC event, compared to a SN~Ia,
the impact effect will be less severe. However, even if there were somewhat less donor material to recycle the
post-AIC pulsars, the main findings in this work would remain intact.

\subsubsection{Super-Chandrasekhar mass WDs}\label{subsubsec:superC}
Recent observations of exceptionally luminous SNe~Ia \citep[e.g.][]{hsn+06,saa+10} suggest that their WD progenitors had a
super-Chandrasekhar mass ($\sim\!2.0-2.5\;M_{\odot}$).
The possibility of super-Chandrasekhar mass WDs has been investigated theoretically for differentially rotating WDs \citep[e.g.][]{yl05},
WDs with extreme magnetic fields \citep[e.g.][]{dm13} and merging WDs \citep[e.g.][]{it84}.\\
Our binary calculations are able to produce such massive WDs, but only in systems with main-sequence star donors.
In the left panel of Fig.~\ref{fig:Mdot_AICNS_12}, we show how WDs up to about $2.1\;M_{\odot}$ (if they exist in nature) can be produced
in these systems (the open circles indicate when the WDs reach a mass of $2.0\;M_{\odot}$). The potential formation of these massive accreting 
WDs was also found in \citet{ldwh00}. It is obvious that the initial parameter space for producing SNe~Ia, or AIC events, 
is much more limited for such super-Chandrasekhar mass WDs.

\subsection{Comparison to previous work}\label{subsec:others}
To explain the existence of high B-field NSs in old stellar populations, or in close binaries in the Galactic disk,   
it has been suggested in the literature that: 1) these NSs are formed via the AIC formation channel (cf. Sect.~\ref{sec:AICobs}), 
and 2) post-AIC accretion, in general, does not affect these newborn NS significantly and therefore they remain
to appear as young pulsars or, at most, mildly recycled pulsars \citep[e.g. ][]{ihr+08}.  
Our calculations partly disagree with the second hypothesis. 
We find that the majority of pulsars formed via AIC would have accreted significant amounts of matter following the AIC event, 
leading to the low B-fields of recycled pulsars (in particular, this is the most likely outcome if the donor is a main-sequence star). 
A similar conclusion was also found by \citet{sl00} who
argue that it is difficult to reproduce high B-field pulsars in close orbits (e.g. PSR~B1831$-$00) via the AIC formation scenario. 
However, as pointed out in Sect.~\ref{subsec:outcome}, 
in a few cases one would indeed expect to find pulsars that are not recycled, or are only very mildly recycled, as a consequence of 
late AIC towards the end of the mass-transfer process, disruption of the post-AIC binary by an encounter in a GC,
or hampering of post-AIC accretion due to the propeller and/or the radio ejection mechanism for systems with giant star or 
helium star donors.
A population synthesis investigation might help to study the probabilities of these events happening.

\subsubsection{The initial ($P_{\rm orb},\,M_2$)--parameter space }\label{subsubsec:initial}
In Fig.~\ref{fig:grid_all} we summarize our results in terms of the initial parameter space of AIC progenitors in a ($P_{\rm orb},\,M_2$)--diagram.
\citet{ldwh00} studied in detail the binary evolution of main-sequence star donors with $0.7-1.0\;M_{\odot}$ WD accretors, somewhat less massive
than the initial WD mass of $1.2\;M_{\odot}$ assumed here.
Three examples of studies which did have a $1.2\;M_{\odot}$ WD accretor include \citet{lv97} and \citet{wlh10}, 
who studied the progenitors of SN~Ia for main-sequence and giant star donors, and \citet{hp04} who studied the SN~Ia progenitors for main-sequence stars. 
One can compare Fig.~2 or Fig.~3 in these three papers to our Fig.~\ref{fig:grid_all}. For giant stars, our results are generally in agreement, 
except for Wang~et~al. who did not reach the Chandrasekhar limit for giant stars in systems with $P_{\rm orb}>25^{\rm d}$, possibly as a consequence
of their inclusion of accretion disk instabilities.  
For the main-sequence donors, however, a key difference is that we are not able to produce systems  
leading to WD collapse for $3\;M_{\odot}$ donor stars. 
In our work, as discussed in Sect.~\ref{subsubsec:MTrates}, we did not allow for the optically thick wind model \citep{ki92,kh94}
to operate for high values of $|\dot{M}_2|$.
With our assumptions, when the mass-transfer rate becomes very high for these massive donors, the fate of the system is a merger. Therefore,
our maximum main-sequence donor star mass is $\sim\!2.7\pm 0.1\;M_{\odot}$. Only if we let $\dot{M}_{\rm CE}=10\;\dot{M}_{\rm Edd,WD}$ 
are we able to produce AIC events with $3\;M_{\odot}$ donor stars.

Whereas \citet{lv97} and \citet{hp04} found that SN~Ia could result from systems having main-sequence donor stars with masses up to $\sim\!3.5\;M_{\odot}$, 
\citet{hkn08} propose solutions with main-sequence donor stars up to $\sim\!7\;M_{\odot}$ by introducing an efficient stripping of the donor star
via impact of a strong wind from the accreting WD. By assuming a stripping rate up to a factor of 10 larger than the wind mass-loss rate of the WD
($\dot{M}_{\rm strip}\la 10\,\dot{M}_{\rm wind}$) the systems can remain dynamically stable and avoid evolving into a CE.  
We find this scenario questionable, both with respect to its realization and its stability,   
although attractive if the main goal is to match the observed SN~Ia rates with theoretical modelling
based solely on the single-degenerate scenario.
In a recent paper \citet{btn13} demonstrated the strong dependence on the
predicted SN~Ia rates on the wind-stripping effect and on the mass accumulation (retention) efficiency of the accreting WD.

\citet{wh10} and \citet{lcwh10} studied the helium star donor channel for SNe~Ia. From their Figs.~2 and 4, respectively, 
we can directly compare their results to ours shown in Fig.~\ref{fig:grid_He}.
The discrepancy is quite large. According to their work, systems with massive helium star donors (up to $\sim\!3\;M_{\odot}$) 
lead to a Chandrasekhar-mass WD, and \citet{wh10} even find that SNe~Ia occur for low-mass helium star donors in very wide systems (up to 100~days). 
We find that only systems with initial helium star donor masses $M_2 \le1.5\;M_{\odot}$ (and $P_{\rm orb}\le 1.2^{\rm d}$) 
are able to produce WDs undergoing AIC (or SN~Ia).
Again, the main reason for this difference is their use of the optically thick wind model, as discussed above. \\
To investigate this question of wind dynamical instability for helium stars donors, we performed two test runs without restrictions from obtained values of $|\dot{M}_2|$.
We considered a $2.0\;M_{\odot}$ and a $2.6\;M_{\odot}$ helium star in a binary with a $1.2\;M_{\odot}$ WD and $P_{\rm orb}=0.10^{\rm d}$.
Although the orbital size shrinks by up to 30~\%, the systems remain dynamically stable in both cases. However, the mass-transfer rates become quite super-Eddington.
For the $2.6\;M_{\odot}$ helium star, $|\dot{M}_2|\simeq 1.0\times 10^{-4}\;M_{\odot}\,{\rm yr}^{-1}$ (corresponding to $|\dot{M}_2| > 100\;\dot{M}_{\rm Edd, WD}$).
Whether or not the optically thick wind model remains valid in this regime is questionable and more research is needed to clarify this important question.

Another factor that determines the parameter space of systems evolving successfully to a critical WD mass (and therefore affects the Galactic SN~Ia birthrate)
is the assumed Chandrasekhar mass. Whereas we applied a threshold mass of $1.48\;M_{\odot}$ (to include the effect of rotation),
a smaller limit of $\sim\!1.38\;M_{\odot}$ was applied by \citet{hp04,wh10} and \citet{wlh10}, which facilitates reaching the critical WD mass point.
 
\subsubsection{Population synthesis studies of MSP formation via AIC }\label{subsubsec:popsyn}
\citet{htw+10} performed a population synthesis study of binary MSPs formed via the AIC channel. A few interesting points become clear when comparing
our results to their Figs.~1d,~2d, and 3d. They produce a large number of very wide systems with CO~WDs and $P_{\rm orb}=10^3-10^6\;{\rm days}$. Although, it is expected
that some systems widen significantly at the moment of the AIC when a small kick is applied, the numbers of these very wide systems is surprising to us. 
The donor stars in these systems are giants which later settle as CO~WDs. However, given that these giants have initial masses $0.9\le M_2/M_{\odot} \le 1.1$, their maximum
radii reach just a few $10^2\;R_{\odot}$ and therefore they are unlikely to serve as donor stars in binaries expanding much beyond 1400~days. Hence, recycled pulsars
are not expected in systems with $P_{\rm orb}>1400\;{\rm days}$ (cf. our models GS1$-$4 and GSZ1). Similarly, we cannot reproduce the distribution
of massive ($\sim\!0.8\;M_{\odot}$) CO~WDs in a wide range of systems with $P_{\rm orb}\simeq 0.1-100\;{\rm days}$. The progenitors of these systems are
binaries with an accreting WD and a helium star donor. We find that for these systems the expected final orbital periods are $P_{\rm orb}\simeq 0.05-2.0\;{\rm days}$. (Even applying
large kicks will not change this picture much.) The reason for these apparent discrepancies is not clear to us and cannot simply be explained by using
different limits for the WD accretion window, although we apply a different prescription for the mass accumulation of helium-rich material.
It is also possible that the BSE~code used by \citet{htw+10} fails to trace the mass-transfer process with sufficient accuracy. 

There are, however, also similarities in our results. Most notably, \citet{htw+10} find that the minimum orbital period of MSPs with He~WD 
companions, produced via AIC, is about 5~days
(we find a minimum $P_{\rm orb}\simeq 10\;{\rm days}$) and that there is a gap in the orbital period distribution between $\sim\!50-200\;{\rm days}$ 
(depending on their chosen model), where 
we find the gap is roughly between $60-500\;{\rm days}$ (cf. Table~\ref{table:AICmodels} for models with no AIC kick, $w=0$).
Ideally, our work in this paper would be used as a basis for a new population synthesis study for comparison with the work of \citet{htw+10}, for example.

\subsection{Space velocities: distinguishing AIC from SNe~Ib/c}
\begin{figure}
\begin{center}
  \includegraphics[width=0.48\textwidth, angle=0]{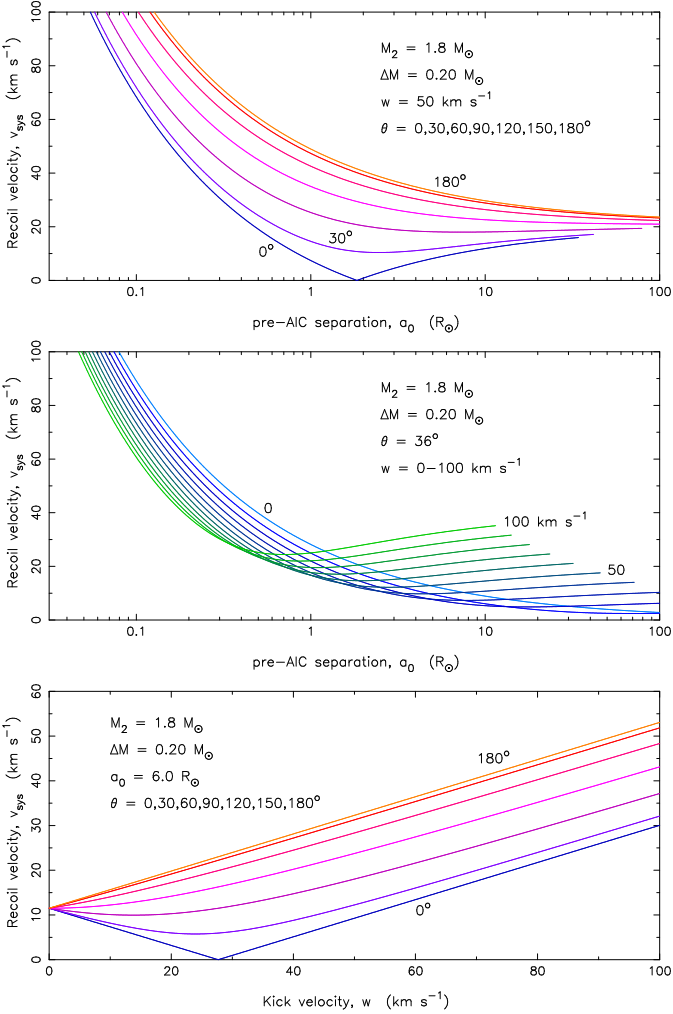}
  \caption[]{Systemic recoil velocities of MSPs formed indirectly via AIC. 
             The two upper panels show sample calculations of the recoil velocities, $v_{\rm sys}$ obtained for a
             fixed kick of $w=50\;{\rm km\,s}^{-1}$ and kick angles $\theta=0-180^{\circ}$ (upper panel), or a fixed value
             of $\theta = 36^{\circ}$ and kicks $w=0-100\;{\rm km\,s}^{-1}$ (central panel), in both cases as a function
             of the pre-AIC orbital separation, $a_0$. The lower panel shows $v_{\rm sys}$ as a function of kick velocities, $w$
             for fixed values of $a_0$ and variable values of $\theta=0-180^{\circ}$. 
	     In all cases we assumed a companion star mass of $M_2^{\rm AIC}=1.8\;M_{\odot}$ at the moment of the AIC.
             AIC events are generally believed to have $w\simeq 0$ (i.e. without a kick). See text for discussions.
    }
\label{fig:recoil}
\end{center}
\end{figure}
In Fig.~\ref{fig:recoil} \citep[partly adapted from ][]{tb96} we show examples of systemic recoil velocities expected for MSPs which formed indirectly via AIC.
For systems with main-sequence donors, the pre-AIC orbital separations are typically $5-10\;R_{\odot}$ and 
the resulting velocities are $v_{\rm sys}\le 50\;{\rm km\,s}^{-1}$, even when applying hypothetical kicks up to $w=100\;{\rm km\,s}^{-1}$. 
For NSs formed via AIC with helium star donors the expected values of $v_{\rm sys}$ are somewhat larger because of their narrower orbits. Neutron stars formed via AIC
with giant star donors only survive in binaries if the associated kicks are very small ($w\ll 50\;{\rm km\,s}^{-1}$) as a result of their large
pre-AIC orbital separations, $a_0>200\;R_{\odot}$.\\
To summarize, we expect binary pulsars formed via AIC to have small
systemic velocities relative to their local stellar environment; as opposed to what is expected for systems where the NS
was formed via standard SN~Ib/c SNe \citep{tb96}. 

\begin{figure}
\begin{center}
  \includegraphics[width=0.35\textwidth, angle=-90]{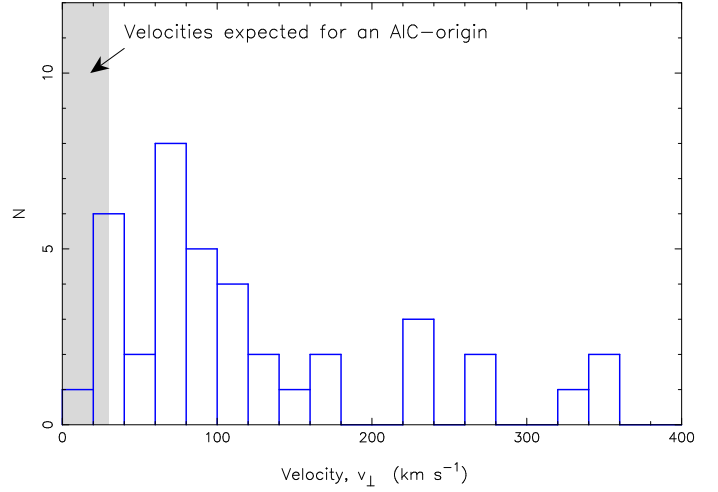}
  \caption[]{All measured transverse velocities of binary radio pulsars with a WD companion.
             Data taken from the {\it ATNF Pulsar Catalogue} in March~2013
             \citep[][http://www.atnf.csiro.au/research/pulsar/psrcat]{mhth05}.
             AIC is expected to form binary pulsars with $v_\perp < 30\;{\rm km}\,{\rm s}^{-1}$. However, the observed
             binary pulsars with such small velocities could also have formed by EC~SNe. 
    }
\label{fig:vel_data}
\end{center}
\end{figure}

In the histogram in Fig.~\ref{fig:vel_data} we have plotted the distribution of all 39 measured transverse velocities, $v_\perp$ 
(derived from proper motions) of binary radio pulsars 
with a WD companion. The average transverse velocity is $\langle v_\perp \rangle = 120\;{\rm km\,s}^{-1}$ and the median value is $89\;{\rm km\,s}^{-1}$.
These values correspond to 3-D systemic space velocities which, in average for random orientation of their velocity vectors with respect to the line-of-sight,
are larger by a factor of $4/\pi$, i.e. $\langle v_{\rm sys} \rangle \approx 160\;{\rm km\,s}^{-1}$. 
(Here we assume that these space velocities correspond to
typical peculiar velocities of the systems with respect to the local standard of rest at their birth locations in the Galactic plane.)
It is seen that the vast majority of the observed systems have much larger $v_{\rm sys}$ than the expected $10-30\;{\rm km\,s}^{-1}$
for NS binaries that were formed via the AIC channel. Based on this alone, one could be tempted to conclude that the fraction of binary pulsars formed
via AIC must be rather small (at least less than 20~\%), if AICs are symmetric or even accompanied by a small kick, $w<50\;{\rm km\,s}^{-1}$.
However, one must keep in mind the selection effects at work given that it is much easier to determine the proper motions of fast moving objects.
In the future when the number of measured velocities has increased significantly it may be possible to identify structured peaks in the 
velocity distribution, reflecting NS origins from AIC (or EC~SNe) and SN~Ib/c. 

\subsection{Formation of high B-field NSs in close Galactic binaries}\label{subsec:highBgal}
In this work, we have presented the suggested observational evidence for pulsars formed via AIC in GCs.
Considering orbital parameters of binary pulsars, however, sources located in GCs are generally  
not suitable as tracers of their binary evolution history because
of the frequent encounters and
exchanges of companion stars in the dense environments \citep{sp93,hhm96,rhs+05}. 
We therefore proceed to discuss only binary parameters of NSs in the Galactic disk.

In Table~\ref{table:AICcandidates} we presented four NS binaries in the Galactic disk
with unusual properties (i.e. relatively slow spins and high B-fields, despite being
in close binaries with an ultra low-mass companion star), see Sect.~\ref{subsec:AICclassic}.
The interesting question now is if we can reproduce these systems with our AIC modelling.  
The answer is no. Whereas the slow spin periods and the large B-fields may be accounted for in some post-AIC binaries
with similar orbital periods less than a few days (see Sect.~\ref{subsubsec:AIC_corbet}), all our calculated 
close-orbit post-AIC systems have massive CO~WD companions with masses $0.65 <  M_{\rm WD}/M_{\odot} < 0.85$
This is in clear contrast to the four binaries in Table~\ref{table:AICcandidates} which all have companion
stars with masses $\le 0.10\;M_{\odot}$. 
One way to reconcile this companion mass discrepancy is if the newborn NSs have evaporated their companion stars
very efficiently after the AIC event. Although the available spin-down luminosity of these slow NSs is relatively small ($\dot{E}\propto P^{-3}$) 
at the present epoch, these NSs could have been much more energetic in the past. An evaporation scenario could possibly also apply to post-AIC systems 
that evolved into (and survived) a CE, in case of main-sequence star donors (cf. Fig.~\ref{fig:PM2-plane} and Sect.~\ref{subsec:PM2-plane}).
The ultra-compact X-ray binary 4U~1626$-$67 exhibits UV and X-ray emission lines of C, O, and Ne which suggests
an evolved helium star or even a WD donor \citep{nyvt10}.
For further discussions on 4U~1626$-$67 and PSR~J1744$-$3922, see \citet{ynv02} and \citet{brr+07}, respectively.
The formation of these four abnormal NS binaries is one of the biggest challenges to our understanding of close binary evolution.

\begin{figure}
\begin{center}
\mbox{\includegraphics[width=0.35\textwidth, angle=-90]{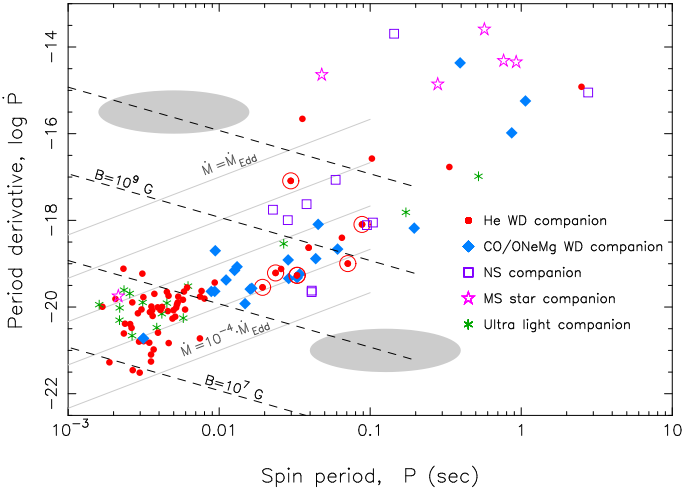}}
  \caption[]{
    The distribution of 115 Galactic binary radio pulsars in the $P\dot{P}$--diagram.
    The error~bars are much smaller than the size of each symbol.
    The caption to the right explains the nature of the companion stars. Pulsars with a He~WD companion marked by a circle 
    are discussed in Sect.~\ref{subsubsec:puzzling}. 
    The B-fields (dashed lines) and the spin-up lines (grey lines) were calculated following
    Tauris, Langer \& Kramer (2012) and assuming $M_{\rm NS}=1.4 M_{\odot}$ and $\sin \alpha = \phi = \omega _{\rm c}=1$.
    Data taken from the {\it ATNF Pulsar Catalogue} in March~2013.  All observed $\dot{P}$ values were corrected for kinematic effects. 
    The grey-shaded areas mark empty regions that cannot be populated according to current recycling scenarios.}
\label{fig:ppdot}
\end{center}
\end{figure}

\subsubsection{Puzzling radio pulsars in the Corbet~diagram}\label{subsubsec:puzzling}
\begin{figure}
\begin{center}
\mbox{\includegraphics[width=0.35\textwidth, angle=-90]{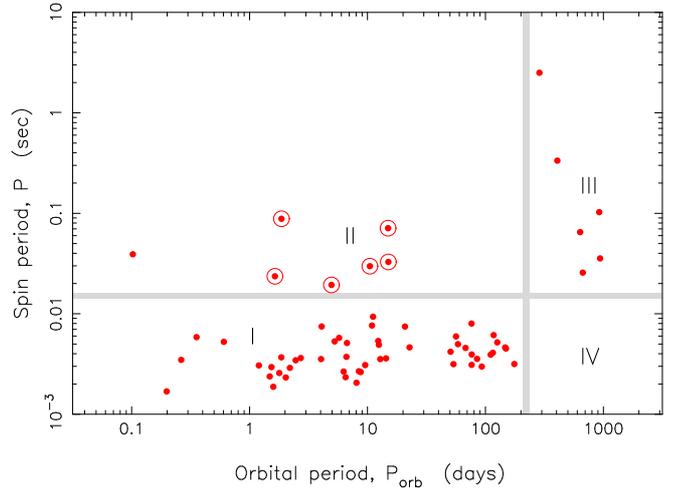}}
  \caption[]{
    The Corbet diagram for the observed distribution of the 64 binary radio
    pulsars with a He~WD companion. The plot reveals four regions, labelled by I, II, III, and IV,
    that may be understood from an evolutionary point of view (see text).
    The pulsars in region~II are also marked by a circle in the $P\dot{P}$--diagram in Fig.~\ref{fig:ppdot}
    and their potential link to an AIC origin is discussed in the text \citep[figure updated after][]{tau11}.
    }
\label{fig:corbet}
\end{center}
\end{figure}

Besides the four systems discussed above,
one may ask if there are additional potential AIC candidates among the binary radio pulsars in the Galactic disk.  
Figure~\ref{fig:ppdot} shows the location of the known Galactic binary pulsars in the $P\dot{P}$--diagram. 
The mildly recycled MSPs ($10\,{\rm ms} < P < 100\,{\rm ms}$) are dominated by systems with CO/ONeMg~WD and NS companions.
As shown in \citet{tlk12}, their relatively slow spin rates are expected from an evolutionary point of view as a consequence of the rapid mass-transfer phase
from a relatively massive donor star.\\
As seen in Fig.~\ref{fig:ppdot} there is also a group of pulsars that have similarly slow spin periods, and relatively large B-fields, 
but with He~WD companions. These pulsars (marked with a large open circle) may also have had a limited recycling 
phase which could hint at a possible alternative origin. Their puzzling nature is clearer when
plotting all binary radio pulsars with He~WDs in the Corbet diagram.
As shown in Fig.~\ref{fig:corbet}, an interesting pattern is revealed and we now briefly discuss four regions in this diagram.

Region~I shows that MSPs can be fully recycled over a spread of 3 orders of magnitude in final orbital period.
If the $P_{\rm orb}\ga 200^{\rm d}$ the pulsars are only partially recycled, 
as noticed from their slow spin periods (region~III). This could be related to the relatively short mass-transfer phase 
in wide-orbit LMXBs where the donor star is highly evolved by the time it fills its Roche~lobe \citep{ts99}. 
In region~II, one sees the small sub-population of puzzling systems (marked by circles and discussed above) 
with $1^{\rm d} < P_{\rm orb} < 200^{\rm d}$, all of which are only mildly
recycled MSPs with spin periods between 20 and 100~ms. Where do these systems come from? And why do they have much
slower spin periods than the pulsars in region~I with similar orbital periods? 
(In this group of marked pulsars in region~II
we exclude PSR~J0348+0432 at $P_{\rm orb}=0.1024\;{\rm days}$ and $P=39\;{\rm ms}$, since this system 
can probably be explained by standard evolution of a converging LMXB system; \citet{afw+13}).

Given their slow spins these six pulsars might originate from progenitor systems  
where the amount of mass accreted was limited. Whereas the populations in regions~I and III are thought to be produced from standard LMXBs 
with different initial orbital periods following a core-collapse SNe, 
the situation might be different for pulsars in region~II.
Could these systems then perhaps \citep[as suggested in][]{tau11} originate from AICs where the subsequent spin~up of the newborn NS
only resulted in a mild spin-up? The rationale is that if a 
limited amount of material remains in the donor star envelope by the time the ONeMg~WD undergoes AIC,  
then only a limited amount of material is available to spin~up the pulsar later on. 
Furthermore, 
the narrow range of orbital periods for these pulsars could then
reflect the finetuned interval of allowed mass-transfer rates needed for the progenitor ONeMg~WDs to accrete and grow in mass to the
Chandrasekhar limit before their implosion.
This is a tempting hypothesis but, as we argue below, we find that this cannot be the case.

In this paper we have demonstrated that MSPs with He~WD companions and $P_{\rm orb}<10^{\rm d}$ do not form via AIC
for the following reason:  
Main~sequence donor stars must have masses above at least $1.6\;M_{\odot}$
to ensure sufficiently high mass-transfer rates for the WD mass to grow to the AIC limit, and therefore 
magnetic braking is not expected to operate in these systems since such massive main-sequence stars do not have convective envelopes.
As a consequence of this, the systems leading to AIC avoid significant loss of orbital angular momentum and their 
final orbital periods will always exceed 10~days. 
Given that three out of the six puzzling pulsars have $P_{\rm orb}< 10\;{\rm days}$ we conclude that the AIC channel cannot explain
the origin of their characteristics. 
Last but not least, two of these six binary pulsars have measured transverse velocities and their values are extraordinarily large:
$v_\perp = 269\;{\rm km\,s}^{-1}$ for PSR~J1745$-$0952 and $v_\perp = 326\;{\rm km\,s}^{-1}$ for PSR~J1810$-$2005,
according to the {\it ATNF Pulsar Catalogue}, based on measurements by \citet{jsb+10}. 
Such large transverse velocities are clearly in contradiction with the expectations of an AIC origin.

\subsubsection{Observed pulsars with CO~WDs in the Corbet~diagram}\label{subsubsec:CorbetCOWD}
For the observed distribution of recycled binary pulsars with CO~WDs in the Corbet diagram, we refer to Fig.~3 in \citet{tlk12}. 
We note that some of our modelled systems with an AIC origin, plotted in Fig.~\ref{fig:AIC_corbet},  
share properties of a subpopulation of the observed systems. These are systems with helium star donors which
leave behind recycled pulsars with $4~{\rm ms} < P < 200~{\rm ms}$ and $P_{\rm orb} < 2^{\rm d}$.
The very wide orbit radio pulsars with CO~WDs, which result from our AIC channel modelling in systems with giant star donors (e.g. model GS3),
also resemble an observed system: PSR~B0820+02. This pulsar is in a 1232~day orbit, has a spin period of 0.86~sec, and a CO~WD companion of 
mass $>0.52\;M_{\odot}$ \citep{kr00}. Hence, it is indeed possible that some of the observed radio pulsars with CO~WDs could have an AIC origin.
If this is the case, then these pulsars should all have a mass close to $1.28\;M_{\odot}$ since they do not accrete much in the post-AIC LMXB phase.
This is in contrast to the binary pulsars with He~WDs, which may accrete $\sim\!0.4\;M_{\odot}$ following the AIC event. 
One observed pulsar with a CO~WD and a low-mass NS is PSR~J1802$-$2124 which has $M_{\rm NS}=1.24\pm0.11\;M_{\odot}$, $P=12.6\;{\rm ms}$, 
$P_{\rm orb}=0.70^{\rm d}$, and a $0.78\pm 0.04\;M_{\odot}$ CO~WD \citep{fsk+10}. This pulsar shares some properties of our AIC models He3 and He4
(cf. Table~\ref{table:AICmodels}).
However, binary pulsars with CO~WDs formed via the standard SN~Ib/c channel are not expected to have accreted much either (since here too 
the mass-transfer phase is short lived, either
in intermediate-mass X-ray binaries with relatively close orbits or LMXBs with very wide orbits and giant star donors). However,
these NSs could have formed in SN~Ib/c with birth masses substantially larger than $1.28\;M_{\odot}$ and hence their measured masses could
be much larger too.

\subsection{Optimal evidence for NSs formed via AIC?}\label{subsec:identifyAIC}
One may ask what observational evidence would be needed to firmly prove the AIC formation channel of NSs? 
This is a difficult question to answer since we do not know the exact NS properties expected from AIC. 
However, we could point to a few hypothetical cases of NSs which are not expected to exist from current
knowledge of pulsar recycling.
For example, it would be interesting if future observational surveys discover either   
a very slowly spinning ($\sim\!100\;{\rm ms}$) radio pulsar associated with a very low B-field ($\sim\!10^8\;{\rm G}$), or
a close binary MSP with a very rapid spin ($<5\;{\rm ms}$) and a high B-field ($>10^{10}\;{\rm G}$), orbiting
a companion star which has experienced mass loss
(see grey-shaded areas in Fig.~\ref{fig:ppdot}).
None of these pulsars is expected to form according to present
recycling and spin-up theory \citep[e.g.][]{tlk12}.
However, even if the second kind of pulsar existed
it would be very unlikely to be detected given that its strong magnetic dipole radiation
would slow down its rapid spin rate within a few $100\;{\rm kyr}$. 
In our view, the best evidence for AIC is the apparently young NSs detected in GCs.


\section{Conclusions}\label{sec:conclusions}

  \begin{enumerate}[I]
      \item We have demonstrated, using a detailed stellar evolution code modelling both the pre- and the post-AIC binary evolution 
            (to our knowledge, for the first time), that MSPs can be formed indirectly via the AIC formation channel.
            In this scenario, a normal NS is formed in an AIC event and which subsequently undergoes recycling
            by accretion (from the same donor star, now in a post-AIC LMXB) leading to the formation of an MSP.
            This scenario is possible for systems with donor stars that are either main-sequence stars, giants stars, or helium stars.
            The first type of donor stars lead to fully recycled MSPs with He~WD companions, whereas the other two types of donors lead to more mildly recycled
            pulsars with mainly CO~WD companions.
            The parameter space of the successful progenitor systems is restricted to three limited areas in the initial ($P_{\rm orb},M_2$)-plane.
      \item Millisecond pulsars formed via AIC are difficult to distinguish from MSPs formed via the standard SN~Ib/c channel with respect to their
            masses, B-fields, spin periods, and WD companions. The reason for this is that the donor stars in most cases
            transfer more than $0.1\;M_{\odot}$ of material to the NS after the AIC~event, in a process which mimics the standard recycling scenario.
      \item Nevertheless, we identify two parameters which can, at least in some cases, be used to differentiate the two formation channels.
            First, MSPs formed via AIC and which have He~WD companions have $P_{\rm orb}$ between 10 and 60~days (or $P_{\rm orb}> 500^{\rm d}$ for giant star donors). 
            Second, the velocities of pulsars formed via AIC are predicted to be less than $30\;{\rm km}\,{\rm s}^{-1}$, which agrees well with
            the hypothesis of AIC being the origin of some pulsars retained in GCs. 
            In contrast, MSPs formed via the standard SN~Ib/c channel and which have He~WD companions may obtain 
            $P_{\rm orb}$ in a continuous wide range between a few~hours and up to $\sim\!1000\;{\rm days}$, and
            their space velocities typically exceed $100\;{\rm km\,s}^{-1}$.
      \item In a few cases where the post-AIC LMXB phase is short-lived, and particularly in GCs where the probability of disruption of a wide binary 
            by an encounter event is large, 
            pulsars formed via AIC will be largely unaffected by, or will avoid, subsequent accretion
            from the donor star. This could, for example, be the explanation for the existence of 
            some GC pulsars which appear to be young.
      \item More research is needed on optically thick wind modelling, and direct observations of SN~Ia remnants to
            constrain pre-SN~Ia wind mass loss, 
            as well as modelling of the prescription for calculating the mass-transfer rates from
            giant star donors with small surface gravities and extended atmospheres.
	    As demonstrated in this paper, the application of the optically thick wind model is very important for the estimated event rate of both 
            AIC and SNe~Ia.
  \end{enumerate}

\begin{acknowledgements}
  It is a pleasure to thank Ed van~den~Heuvel for very helpful references and comments on the manuscript.
  T.M.T. gratefully acknowledges support and hospitality from the 
  Argelander-Insitut f\"ur Astronomie, Universit\"at Bonn
  and the Max-Planck-Institut f\"ur Radioastronomie.
\end{acknowledgements}

\bibliographystyle{aa}
\bibliography{tauris_refs}
\newpage
%
\begin{sidewaystable*}
\begin{minipage}[t][ 0mm]{\textwidth}
\caption{Summary of selected systems which evolved successfully to AIC and later produced recycled pulsars 
(see first paragraph of Sect.~\ref{sec:AICMS} for explanations of all the variables). \newline
$\qquad$ [$^*=Y_c$, $\;\;^{**}=$ (hybrid)~CO~WD, $\;\;^{***}=$ CO~WD after He-flash on the RGB, $\;\;$CE $=$ onset of a common envelope].}
\centering
\begin{tabular}{lccrlcccrlrlcrlcccrr}
\hline\hline             
\noalign{\smallskip}
Model & $M_2^{\rm \,ZAMS}$     &   $P_{\rm orb}^{\rm \, ZAMS}$   &  $t_{\rm RLO}$  & $X_c$ & $\Delta t_{\rm CV}$      &   $M_2^{\rm \, AIC}$        &
$P_{\rm orb}^{\rm \, AIC}$     &   kick$^*$         &   angles   &  $v_{\rm sys}$     &  $\Delta t_{\rm detach}$ & $P_{\rm orb}^{\rm \, circ}$ &
$\Delta t_{\rm LMXB}$          &   $M_{\rm WD}$                  &    $P_{\rm orb}^{\rm \, MSP}$                 &
$M_{\rm NS}$                   &   $\Delta M_{\rm NS}$           &    $P_{\rm spin}$                              &   $t_{\rm total}$ \\ 
\noalign{\smallskip}
\noalign{\smallskip}
      & $M_{\odot}$            &   days                          &    Myr          &       & Myr                  &  $M_{\odot}$          & 
days                           &   km$\,$s$^{-1}$   &   deg.     &    km$\,$s$^{-1}$       & Myr                  & days                  & 
Myr                            &   $M_{\odot}$                   &    days                                        &
$M_{\odot}$                    &   $M_{\odot}$                   &    ms                                          &  Myr \\
\noalign{\smallskip}
\hline
\hline
\noalign{\smallskip}
MS1W &  3.00 & 1.50 &  294 & 0.12      & 1.27 & 1.65 & 0.78 & 0   &                &   12 & 0.054    & 0.88 & 335  & 0.310        & 17.8  &  1.46 & 0.18 & 1.2 & 630\\
\hline
\noalign{\smallskip}
MS2W &  2.80 & 5.00 & 1250 & 0.00      & 0.51 & 1.56 & 3.19 & 0   &                &    8 & 0.064    & 3.65 & 4    & 0.411$^{**}$ & 37.2  &  1.31 & 0.03 & 4.7 & 1250\\
\hline
\hline
\noalign{\smallskip}
MS3  &  2.60 & 1.00 &  357 & 0.29      & 1.90 & 1.99 & 0.68 & 0   &                &   13 & 0.003    & 0.76 & 1240 & 0.262        & 23.8  &  1.55 & 0.27 & 0.9 & 1600\\
MS4  &       &      &      &           &      &      &      & 50  & (0,$\,$0)      &    7 & 0.097    & 1.12 & 891  & 0.273        & 31.8  &  1.61 & 0.33 & 0.8 & 1250\\
MS5  &       &      &      &           &      &      &      & 450 & (107,$\,$90)   &  180 & 165      & 1.83 & 214  & 0.298        & 43.6  &  1.66 & 0.38 & 0.7 & 736\\
\hline
\noalign{\smallskip}
MS6  &  2.50 & 1.00 & 403  & 0.28      & 1.98 & 1.99 & 0.72 & 0   &                &  13  & 0.003    & 0.80 & 1180 & 0.264        & 24.8  &  1.56 & 0.28 & 0.9 & 1590\\
\noalign{\smallskip}
MS7  &  2.50 & 1.50 & 504  & 0.08      & 1.68 & 1.97 & 1.07 & 0   &                &  11  & 0.010    & 1.20 & 252  & 0.301        & 26.4  &  1.56 & 0.28 & 0.9 & 758\\
MS8  &       &      &      &           &      &      &      & 50  & (0,$\,$0)      &  9   & 0.19     & 1.87 & 159  & 0.307        & 41.0  &  1.64 & 0.36 & 0.7 & 665\\
MS9  &       &      &      &           &      &      &      & 450 & (116,$\,$90)   &  183 & 58       & 2.88 & 30   & 0.320        & 52.7  &  1.44 & 0.16 & 1.3 & 594\\
\noalign{\smallskip}
MS10 &  2.50 & 2.50 & 539  & 0.00      & 0.67 & 1.88 & 1.76 & 0   &                &  9   & 0.048    & 1.99 & 12   & 0.335$^{**}$ & 33.4  &  1.35 & 0.07 & 2.5 & 552\\
\noalign{\smallskip}
MS11 &  2.50 & 3.00 & 545  & 0.00      & 0.63 & 1.92 & 2.12 & 0   &                &  9   & 0.124    & 2.39 & 11   & 0.341$^{**}$ & 38.3  &  1.34 & 0.06 & 2.8 & 557\\
\hline
\noalign{\smallskip}
MS12 &  2.20 & 1.00 & 584  & 0.26      & 2.08 & 1.85 & 0.79 & 0   &                &  12  & 0.016    & 0.89 & 986  & 0.269        & 27.3  &  1.60 & 0.32 & 0.8 & 1570\\
MS13 &       &      &      &           &      &      &      & 50  & (0,$\,$0)      &  8   & 100      & 1.34 & 450  & 0.285        & 37.4  &  1.67 & 0.39 & 0.7 & 1140\\
MS14 &       &      &      &           &      &      &      & 450 & (112,$\,$90)   &  189 & 264      & 2.02 & 108  & 0.297        & 47.3  &  1.58 & 0.30 & 0.8 & 958\\
\hline
\noalign{\smallskip}
MS15 &  2.20 & 2.00 & 765  & 0.00      & 1.05 & 1.84 & 1.58 & 0   &                &  10  & 0.046    & 1.79 & 29   & 0.295        & 40.1  &  1.42 & 0.14 & 1.5 & 795\\
\noalign{\smallskip}
MS16 &  2.20 & 2.50 & 770  & 0.00      & 0.91 & 1.88 & 1.98 & 0   &                &  9   & 0.178    & 2.24 & 25   & 0.303        & 48.1  &  1.40 & 0.12 & 1.7 & 796\\
\hline
\noalign{\smallskip}
MS17 &  2.00 & 1.00 & 788  & 0.24      & 4.92 & 1.59 & 0.84 & 0   &                &  12  & 0.487    & 0.96 & 839  & 0.272        & 28.3  &  1.59 & 0.31 & 0.8 & 1630\\
\noalign{\smallskip}
MS18 &  2.00 & 2.00 & 1010 & 0.00      & 2.02 & 1.63 & 1.68 & 0   &                &  9   & 6.74     & 1.92 & 66   & 0.297        & 41.8  &  1.44 & 0.16 & 1.3 & 1090\\
\hline
\hline
\noalign{\smallskip}
MZ1  &  2.00 & 1.00 & 594  & 0.00      & 1.00 & 1.30 & 0.99 & 0   &                &   11 & 0.003    & 1.16 & 22.5 & 0.310        & 20.5  &  1.36 & 0.08 & 2.3 & 618\\
\hline
\noalign{\smallskip}
MZ2  &  1.80 & 4.00 & 811  & 0.00      & 1.95 & 1.44 & 3.62 & 0   &                &  7   & 1.62     & 4.17 & 56   & 0.345        & 58.6  &  1.44 & 0.16 & 1.3 & 871\\
MZ3  &       &      &      &           &      &      &      & 50  & (0,$\,$0)      &  16  & 15.0     & 8.19 & 47   & 0.367        & 95.3  &  1.44 & 0.16 & 1.3 & 875\\
MZ4  &       &      &      &           &      &      &      & 450 & (161,$\,$90)   &  218 & 21.4     & 10.57& 41   & 0.377        & 114.7 &  1.44 & 0.16 & 1.3 & 875\\
\hline
\noalign{\smallskip}
MZ5  &  1.40 & 1.00 & 1750 & 0.00      & 4.52 & 0.957& 1.26 & 0   &                &    9 & 0.228    & 1.51 & 165  & 0.285        & 21.7  &  1.45 & 0.17 & 1.3 & 1920\\
\hline
\hline
\noalign{\smallskip}
GS1  &  1.10 & 80   & 8510 & 0.00      & 2.29 & 0.711& 134  & 0   &                &  2   & 3.83     & 163  & 5.01 & 0.408        & 552   &  1.31 & 0.02 & 6.4 & 8520\\
\noalign{\smallskip}
GS2  &  1.10 & 180  & 8520 & 0.00      & 1.69 & 0.685& 325  & 0   &                &  1   & 1.91     & 394  & 2.40 & $\sim\!0.47^{***}$&$>919$ &  1.29 & 0.011& 14  & 8530\\
\hline
\noalign{\smallskip}
GS3  &  1.00 & 400  &13600 & 0.00      & 1.98 & 0.577&  907 & 0   &                &  1   & 2.51     & 1110 & $\sim\!0.01$& $\sim\!0.50^{***}$&$\sim\!1400$&1.280&0.000& 500  & $13600$\\
\hline
\noalign{\smallskip}
GS4  &  0.90 & 160  &19100 & 0.00      & 3.63 & 0.529& 372  & 0   &                &  1   & 0.80     & 459  & CE   & $\sim\!0.46^{**}$ & CE &  1.280& 0.000& --  & --\\
\hline
\noalign{\smallskip}
GSZ1 &  1.00 & 250  & 6330 & 0.00      & 2.02 & 0.622& 481  & 0   &                &  1   & 0.30     & 587  & 2.32 & $\sim\!0.53^{***}$ & $>819$ &  1.32 & 0.010 & 11.0 & 6340 \\
\hline
\hline
\noalign{\smallskip}
He1  &  1.40 & 0.04 & 0.014& 0.97$^*$  & 2.77 & 1.05 & 0.039& 0   &                &   30 & 3.59     & 0.046& 7.96 & 0.670$^{**}$ & 0.072 &  1.32 & 0.04 & 3.8 & 14.3\\
\noalign{\smallskip}
He2  &  1.40 & 0.05 & 2.49 & 0.51$^*$  & 5.87 & 1.03 & 0.044& 0   &                &   29 & 0.073    & 0.051& 1.32 & 0.719$^{**}$ & 0.097 &  1.294& 0.014& 8.4 & 9.75\\
\noalign{\smallskip}
He3  &  1.40 & 0.10 & 7.12 & 0.00$^*$  & 0.20 & 1.12 & 0.100& 0   &                &   23 & 0.043    & 0.118& 0.392& 0.821$^{**}$ & 0.199 &  1.284& 0.004& 21  & 7.75\\
\hline
\noalign{\smallskip}
He4  &  1.20 & 0.05 & 4.26 & 0.00$^*$  & 5.60 & 0.797& 0.054& 0   &                &   25 & 0.256    & 0.066& 1.07 & 0.703$^{**}$ & 0.082 &  1.291& 0.011&  10 & 11.2\\
\noalign{\smallskip}
He5  &  1.20 & 0.60 & 21.9 & 0.00$^*$  & 0.11 & 0.921& 0.707& 0   &                &   11 & 0.005    & 0.84 & 0.121& 0.833$^{**}$ &  1.03 &  1.281& 0.001&  60 & 22.1\\
He6  &       &      &      &           &      &      &      & 50  & (0,$\,$0)      &   18 & 0.017    & 1.30 & 0.113& 0.836$^{**}$ &  1.53 &  1.281& 0.001&  60 & 22.1\\
He7  &       &      &      &           &      &      &      & 450 & (120,$\,$90)   &  267 & 0.023    & 1.66 & 0.108& 0.838$^{**}$ &  1.95 &  1.281& 0.001&  60 & 22.1\\
\hline
\noalign{\smallskip}
He8  &  1.10 & 1.20 & 32.4 & 0.00$^*$  & 0.16 & 0.829& 1.52 & 0   &                &    8 & 0.012    & 1.81 & 0.032& 0.826$^{**}$ &  1.82 &  1.280& 0.000& 200 & 32.6\\
\hline
\hline
\label{table:AICmodels}
\end{tabular}
\end{minipage}
\end{sidewaystable*}

\end{document}